\documentclass{aa}  
\usepackage[varg]{txfonts}
\usepackage{siunitx}
\usepackage{nccmath}
\usepackage{url}
\usepackage[normalem]{ulem}
\usepackage{multirow}
\usepackage{booktabs}
\DeclareSIUnit\parsec{pc}  
\DeclareSIUnit\year{yr}
\DeclareSIUnit\erg{erg}

\makeatletter
\renewcommand*\aa@pageof{, page \thepage{} of \pageref*{LastPage}}

\makeatother

\usepackage{graphicx}
\usepackage[colorlinks=true, citecolor=blue]{hyperref}
\usepackage{cleveref}
\usepackage{makecell}
\usepackage{changes}

\definechangesauthor[name=Dieter, color=magenta]{DH}
\begin{document} 
\title{Measurement of femtogauss intergalactic magnetic fields\\ towards \object{Mkn 501}}

\author{Matías Sotomayor Webar \and Yosef Abed \and Dieter Horns}
\institute{Institut für Experimentalphysik, University of Hamburg, Luruper Chaussee 149, D-22761 Hamburg, Germany\\
\email{matias.sotomayor.webar@uni-hamburg.de}}

\authorrunning{Sotomayor Webar et al.}
\titlerunning{Intergalactic Magnetic Field}

\date{Received xxxxx; accepted }

\abstract
{The intergalactic magnetic field (IGMF) is expected either to be generated  via astrophysical processes or to be a relic of phase transitions in the early universe. Upper bounds  are set via observations of Faraday rotation measure. Lower bounds have been derived from the non-detection of secondary gamma-rays possibly produced in electromagnetic cascades.}
{We investigate the presence of IGMFs by studying the GeV gamma-ray emission from the nearby blazar Mkn~501 ($z=0.034$), searching for evidence of the extended halo from electromagnetic cascades.}
{We analyse 14 years of data from \textit{Fermi}-LAT and \textit{Swift}-XRT/BAT to construct a time-average synchrotron-self Compton model for the TeV spectrum of Mkn~501. This injection spectrum is used to simulate the resulting cascade emission with the ELMAG code for different magnetic field and coherence length configurations. These templates are fit to the \textit{Fermi}-LAT data to find a best-fitting model for the cascade emission.}
{We find significant ($\ge 5 \sigma$ trial-corrected) evidence of extended secondary emission around Mkn~501, which is consistent with an IGMF with 
$B_\mathrm{rms}=1.5^{+2.0}_{-1.3}~\mathrm{fG}$
and a coherence length of $\ell_C=10^{+490}_{-5} ~\mathrm{kpc}$ (statistical uncertainties only). When including instrumental systematic uncertainties, the permissible IGMF strength is $B_\mathrm{rms}=1.5^{+8.5}_{-1.1}~\mathrm{fG}$, for fixed $\ell_C=10~\mathrm{kpc}$. The source needs to actively inject TeV gamma-rays for at least \num{45000} years to match the level of secondary emission.}
{Our results indicate that the secondary gamma-rays are significantly present in the $\it{Fermi}$-LAT data and furthermore, 
that the effect of plasma-heating by pairs in the cascade appears to be negligible for Mkn~501.  The energy spectrum of Mkn~501 favours pair conversion in the intervening low-density
inter-galactic medium.}

\keywords{Astroparticle physics - Magnetic fields - BL Lacertae objects: individual: Mkn 501 - Intergalactic medium - Gamma rays: general}

\maketitle
%
\section{Introduction}

Magnetic fields of the order of 10 nG have recently been detected in cosmic filaments through Faraday rotation measures (RM) \citep{ 2023SciA....9E7233V,2025A&A...693A.208C}. Additionally, $\mu$G fields are known to exist in galaxies and galaxy clusters and have been measured using RM \citep[see e.g.][]{2002ARA&A..40..319C} and confirmed via synchrotron emission under assumption of equipartition between energy densities in particles and fields \citep{2005AN....326..414B}.
Conversely, the existence of an intergalactic magnetic field (IGMF) that has acted as seed for magnetic fields in higher density regions has not been established through observations. So far, only upper and lower limits for the field strength have been obtained through RM and gamma-ray propagation studies, respectively. The coherence length is bound to be smaller than the Hubble scale and larger than theoretically motivated dissipation scales.

The IGMF could be a relic of processes in the early universe, where magnetic fields are suspected to be generated as early as during the inflationary phase, \citep{Turner:1987bw, 1992ApJ...391L...1R}, Electroweak phase transition \citep{VACHASPATI1991258}, Quark-hadron transition \citep{Cheng:1994yr,1989ApJ...344L..49Q}, or in the era of recombination \citep{2013PhRvL.111e1303N}.
In addition to processes related to the early universe, IGMFs could be generated through processes such as the Biermann battery \citep{PhysRev.82.863} during the phase of re-ionization \citep{2000ApJ...539..505G, 2015MNRAS.453..345D, 2025arXiv250421082C} or at later stages in galaxies and propagated into the medium via outflows \citep{2006MNRAS.370..319B}. 

The most sensitive probe for weak IGMF relates to the deflection of pairs produced by TeV-photons injected by emission regions located in the relativistic outflow of active galactic nuclei or gamma-ray bursts. 
The deflected pairs lose their energy predominantly via inverse Compton scattering on low-energy photons of the cosmic-microwave background. 
With an increasing magnetic field, the secondary emission is delayed \citep{1995Natur.374..430P} and appears as an extended halo at GeV energies that can be used to measure or constrain the intervening magnetic field \citep{2009PhRvD..80l3012N}.
The lower bounds on the IGMF \citep[see e.g.][]{2010Sci...328...73N,
2011A&A...529A.144T,2011ApJ...727L...4D,2023ApJ...950L..16A, 2023A&A...670A.145A} rely on the assumption that plasma heating processes by the pairs are a sub-dominant energy-loss mechanism. 

This assumption has been disputed by theoretical investigations \citep{2012ApJ...752...22B,2012ApJ...758..102S}. Recent particle-in-cell simulations indicate the importance of the energy spectrum assumed for the pairs injected \citep{2018ApJ...857...43V}. 
Plasma heating could be suppressed by the interaction with the background plasma \citep{2013ApJ...770...54M}. It is generally accepted that with increasing luminosity of the source the effect of plasma heating increases such that the current lower limits would be relaxed wherever the conditions favour plasma heating over radiative energy losses. Recent theoretical 
studies indicate that the presence of a sizeable magnetic field in the intervening medium suppresses the non-radiative energy losses \citep{2022ApJ...929...67A}.

In this paper, we study the GeV gamma-ray emission of the nearby BL-Lac object Mkn~501 ($z=0.034$), where we search for evidence of the extended secondary emission produced in the pair cascade by performing a likelihood ratio analysis. The paper is structured as follows, in Section 2, we describe the datasets and reduction steps, in Section 3, the data preparation and source model are presented, Section 4 details the construction of our templates from simulations, Section 5 includes our likelihood analysis, and Section 6 contains the results followed by Section 7, where we summarize and present a discussion of our results. Additional material related to the simulation of different source duty cycles, time-delay of secondary emission and systematic effects related to the point-spread function are given in Appendices \ref{app:duty-cycle}, \ref{app:time-delay}, and \ref{app:PSF-bracketing} respectively.
\begin{figure*}[!ht]
    \centering
    \includegraphics[trim=130 10 0 45,clip, scale=0.39]{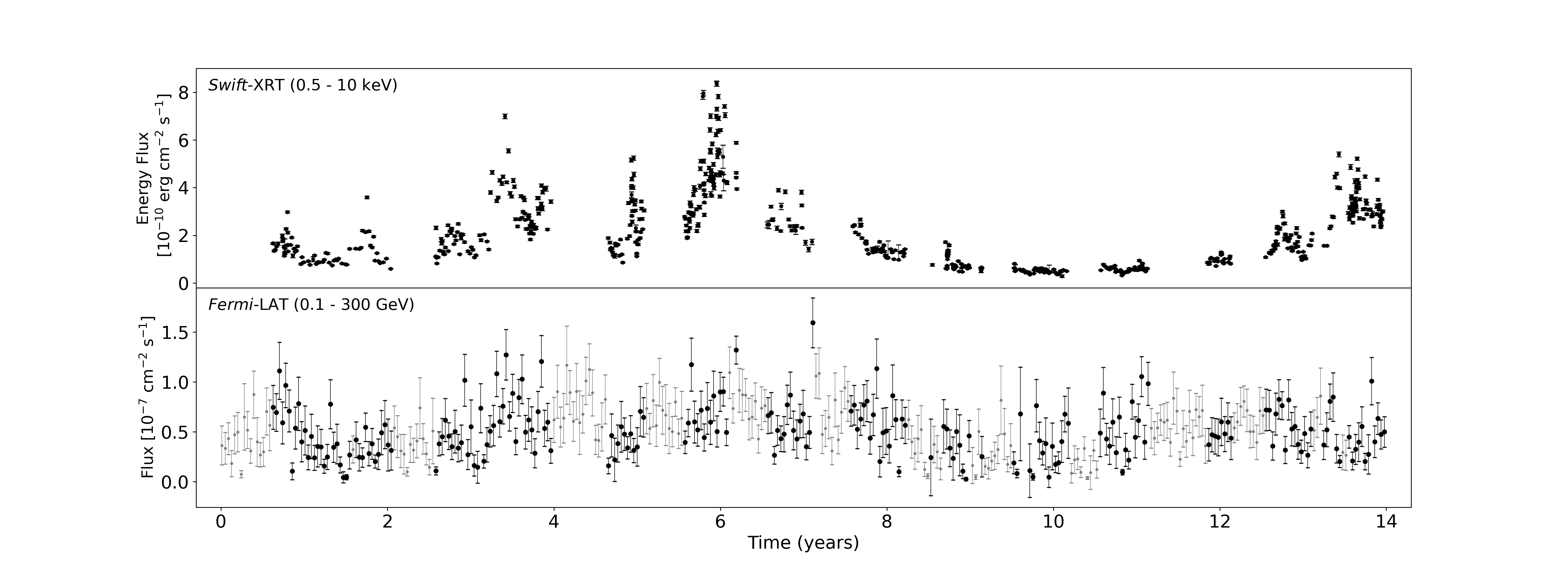}
    \caption{ The light curves of Mkn~501 from August 2008 until August 2022 at X-ray energies (\textit{Swift}-XRT, upper panel: energy flux from \SIrange{0.5}{10}{\kilo\eV}) and at 
    gamma-ray energies (\textit{Fermi}-LAT, lower panel: integrated photon flux from \SIrange{0.1}{300}{\giga\eV}). Each point of the X-ray light curve represents a single pointed observation while the \textit{Fermi}-LAT data are integrated over 2~weeks of observation. Black points indicate the intervals with contemporaneous coverage at X-rays.}
    \label{fig:mkn501_LC}
    
\end{figure*}
\section{Observations of Mkn~501 at X-ray and gamma-ray energies}
\label{sec:gamma-data-model}
In the following, we are estimating a representative energy spectrum extending to several tens of TeV by
reconstructing a spectral energy distribution from contemporaneous X-ray  and gamma-ray observations in conjunction with 
a single-zone synchrotron-self Compton emission model. 
\subsection{X-ray observations with Swift-XRT and BAT}
 The Neil Gehrels Swift observatory was taking data on Mkn~501 in pointed mode with the X-ray telescope (XRT) from 2008 to 2022. In total, 1014 observations were accumulated in windowed-timing mode with a usable observation time of at least 300~s.  
 These XRT data are reduced and processed using the \texttt{xrtpipeline} tool included in the HEASOFT package (v6.34)\footnote{\citet{2014ascl.soft08004N}}. This tool performs a sequence of tasks that prepare the X-ray data for further analysis, such as selecting good time intervals, filtering, creating response files, and binning the data in time and pulse-height.

The wide-field hard X-ray imager (\textit{burst alert telescope}: BAT) onboard the \textit{Swift} satellite is continuously recording data in so-called survey mode. 
The survey data taken during XRT pointed observations from Mkn~501 are selected and reduced using the \texttt{BatAnalysis} package (v2.0.2) \citep{2023zndo...7916509P,2023ApJ...953..155P}. 
Data from 2008--2022 are combined in mosaics for further analysis to extend the energy coverage to \SIrange{15}{100}{\kilo\eV}. 
 
 \subsection{Gamma-ray observations with Fermi-LAT}
 \label{sec:gamma-obs}
Gamma-ray data were recorded with the \textit{Fermi}-LAT \citep{2009ApJ...697.1071A}, and are selected from 14 years of operation (2008--2022) which is the basis of the fourth data-release (4FGL-DR4) of the Fermi catalogue \citep{2023arXiv230712546B}.
We define a region of interest (ROI) of 10$^\circ \times 10^\circ$  centred on the position of the nearby TeV-blazar Mkn 501 ($z=0.034$).
The ROI is subdivided in 500$\times$500 spatial bins with a side-length of $0.02^\circ$.
The energy range between 0.1~GeV and 300~GeV is divided in 28 equal-sized logarithmic bins, corresponding to 8 bins per decade.

The data are filtered following standard practices for analysis of point-like and moderately extended sources\footnote{(\texttt{'DATA$\_$QUAL>0 $\&\&$ LAT$\_$CONFIG==1'}), \texttt{evclass==128}, \texttt{evtype==3}}.
For the instrument response function (IRF), we use the most recent release (\texttt{P8R3$\_$SOURCE$\_$V3}) that is consistent with the event-filtering applied. The data are reduced using the \texttt{Fermitools-conda} (v2.2.0)\footnote{\url{https://github.com/fermi-lat/Fermitools-conda/wiki/Release-Notes\#220-2022-06-21}}. The high-level analysis related to light curves and spectral energy distribution (SED) are carried out with \texttt{fermipy} (v1.2)\footnote{\url{https://github.com/fermiPy/fermipy/releases/tag/v1.2}}.  

\section{Light curves and energy-spectra from contemporaneous observations}
\label{sec:contemporaneous-obs}
\subsection{X-ray light curve}
 
The 1014 X-ray spectra obtained from XRT pointed observations are fit individually with an absorbed power-law (\texttt{cflux*phabs*power}) using \texttt{xspec}. The column density of neutral hydrogen gas is fixed to the Galactic value along the line of sight towards Mkn~501: $n_H=\SI{1.7e20}{\per\square\cm}$ from \citep{2005A&A...440..775K,2016A&A...594A.116H}.
The best-fitting forward-folded integrated energy flux between \SI{0.5}{\kilo\eV} and \SI{10}{\kilo\eV} is calculated with its $68~\%$ uncertainty interval. 
The resulting light curve of energy-flux measurements is shown in Fig.~\ref{fig:mkn501_LC} (upper panel). 
The observations sample the flux state over the entire 14-year period with some gaps in between the monitoring campaigns. The light curve includes episodes of increased variability amplitude in the years 2010/2011, 2012, 2013 followed by a continuous decline with little variability until 2019. Subsequently, the flux gradually increases until episodes of increased variability are observed again in 2021--2022.  

\subsection{Gamma-ray light curve}
The gamma-ray flux light curve is calculated using the \texttt{lightcurve} method implemented in \texttt{fermipy}, integrating the flux over a period of two weeks for each data point. 
An individual likelihood fit is performed for each time bin, where the normalisation for the photon flux of Mkn~501 is kept free. The other parameters related to the sources in the ROI are fixed to the values obtained from a fit to the entire data set. 

The photon flux integrated from \SIrange{0.1}{300}{\giga\eV} is extracted from each fit and plotted in the lower panel of Figure \ref{fig:mkn501_LC}. Those data points that have contemporaneous X-ray data are shown in black.

Overall, the gamma-ray flux appears to follow the X-ray variability. However, the relative uncertainty on the gamma-ray flux measurements is considerably larger than that for the X-ray flux, making a direct comparison more challenging and beyond the scope and intention of this paper.

\subsection{Multi-wavelength spectral energy distribution}
\label{section:MWL}
The background-subtracted 1014 X-ray photon-count spectra from the XRT observations are averaged using the exposure as weights. In the same way, a time-averaged response file is computed. Similarly, the BAT mosaics accumulated for intervals of 2 years are averaged and fit to a power-law model spectrum.

The combined unfolded and rebinned X-ray spectral energy distribution from 0.5~keV to 100~keV is shown in Fig. \ref{fig:SSC-fit-501}. The X-ray spectrum shows a spectral softening across the energy range covered. 
The flat X-ray spectrum in the XRT energy range indicates that the broad peak of the SED is located at keV-energies. 
In the standard paradigm for TeV blazars, the X-rays are believed to be produced via synchrotron radiation from energetic electrons. In the synchrotron self-Compton scenario, the same population of electrons up-scatters the synchrotron radiation via inverse Compton processes to gamma-rays, forming a second peak in the spectral energy distribution.

Since the population of electrons varies over time, the observations at different energies should be strictly simultaneous. Here, we reconstruct a gamma-ray energy spectrum that is contemporaneous to the X-ray observations. This is achieved by taking the central time from each XRT pointed observation, and extracting the $\it{Fermi}$-LAT exposure within $\pm$ 12 hours of the XRT pointing. 
The gamma-ray SED is computed using the \texttt{SED} method implemented in \texttt{fermipy}. The flux normalisation for Mkn~501 is determined for each energy bin separately, leaving the remaining background parameters fixed.

The resulting contemporaneous gamma-ray SED is presented in Fig. \ref{fig:SSC-fit-501}. 
The data points follow a curved SED, rising with energy such that the peak of the inverse Compton emission is located at several 100 GeV or beyond. 

For comparison and historical context, we include ground-based IACT observations from MAGIC and VERITAS \citep{2011ApJ...727..129A} taken during a pre-planned observational campaign in 2009 and \textit{Swift}-UVOT data taken from \citep{2024A&A...685A.117M}. Since both campaigns were not triggered by a high flux state, they are most likely representative for the average flux state of the source given its small duty cycle of flares. 
We include data from HEGRA observations of a flaring state of the source \citep{1999A&A...349...11A} to illustrate the large variability of the TeV flux.
The MAGIC, VERITAS and HEGRA flux points were corrected for absorption in the extragalactic background light (EBL)\footnote{We computed the EBL absorption using the \texttt{ebltable} python package \citep{ebltable_2022}} using the model by \citet{2010ApJ...712..238F}.

\subsection{Time-averaged TeV spectrum from SSC fitting}
\label{sec:TeV-SSC}
\begin{figure}
\centering
\includegraphics[trim=0 0 0 0,clip, scale=0.46]{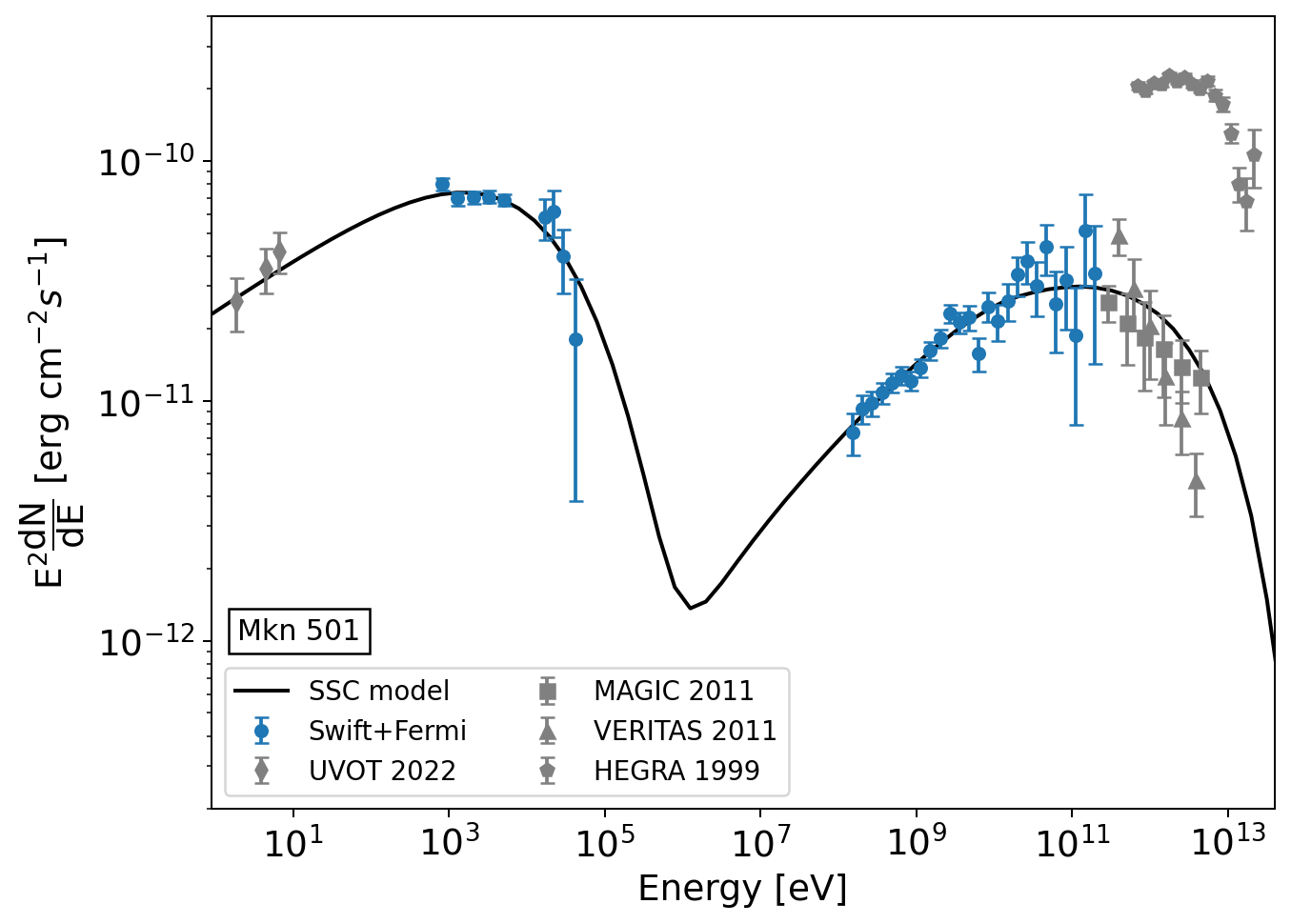}
\caption{The time-averaged spectral energy distribution for Mkn~501 reconstructed using \textit{Swift}-XRT and BAT observations contemporaneous with \textit{Fermi}-LAT data from 2008--2022. Archival data from UVOT, MAGIC, VERITAS and flaring-state from HEGRA are included as reference together with the best-fitting single-zone Synchrotron Self-Compton model (see Sect. \ref{sec:TeV-SSC} for further details).}
\label{fig:SSC-fit-501}
\end{figure}

Currently, there is no representative measurement of the TeV energy spectrum available, even though the FACT collaboration has been monitoring the source extensively in the recent years \citep{2021A&A...655A..93A}. The archival TeV data presented in Fig.~\ref{fig:SSC-fit-501} cover only a five month period in 2009, which corresponds to a rather low flux in comparison to the time-average. Consequently, the archival measurements would under-estimate the injected power at TeV energies averaged over the longer period covered here. 

We use a model-fit to the contemporaneous SED presented in Section~\ref{section:MWL} to estimate the long-term average. 
Long-term optical monitoring of the TeV blazar PKS~2155-304 reveals a break from red- to white-noise behaviour on a characteristic time-scale of $2.7^{+2.2}_{-1.6}$~yr \citep{2011A&A...531A.123K}, indicating that stochastic flux variations decorrelate on longer time-scales. Although this result concerns a different source and waveband, it motivates the use of the $\approx15$-yr averaged SED as  representative for the entire duty cycle of the source.

These data are combined and converted into a \textit{dataset} object that can be used in the framework of the \texttt{gammapy} library \citep{2023A&A...678A.157D}. 

The model SED is calculated using the \texttt{agnpy} package \citep{2022A&A...660A..18N} that models the synchrotron self-Compton emission from electrons in a spherical volume that moves along the jet axis of the blazar. The plasma in the spherical volume is magnetised with a random field with $B_{\mathrm{SSC}}$ and moves with the velocity $\beta$ and bulk Lorentz factor $\Gamma=(1-\beta^2)^{-1/2}$ along a direction inclined by an angle $\vartheta$ with respect to the line of sight, such that the resulting Doppler boost factor $\delta$ is given by:

\begin{ceqn}
\[
\delta = \frac{1}{\Gamma (1-\beta\cos\vartheta)}
\]
\end{ceqn}
The comoving radius $R'$ of the emitting volume is limited by the apparent time-scale of variability $t_\mathrm{var}$, such that
\begin{ceqn}
\[
R'(1+z)\le\delta ~t_\mathrm{var}~c
\]
\end{ceqn}

The spectral distribution of the electron populations within the emitting volume is modelled using a power-law of the electron's Lorentz factor $\gamma$ with normalisation $K$ and power-law index $P$, multiplied by an exponential cut-off function with a cut-off at $\gamma_c$:
$$
\label{eq:PLEC_eq_EED}
\mathrm{n}(\gamma) = K \gamma^{P}~\exp{\left( -\frac{\gamma}{\gamma_c} \right) H(\gamma-\gamma_\mathrm{min})H(\gamma_\mathrm{max}-\gamma)},
$$
with $H(x)=1$ for $x\ge 0$ and $H(x)=0$ for $x<0$. 

The best-fitting parameters are estimated using a $\chi^2$-minimization of the model to match the data. The binning of the spectral flux points is comparable to or even wider than the energy resolution such that the off-diagonal elements in the covariance matrix can be neglected in the fitting procedure.
In Table \ref{tab:SSC-pars} we present the resulting best-fitting values for the model parameters that were left to vary: $K$, $\gamma_c$, $P$, and $B_\mathrm{SSC}$. The values for $\gamma_\mathrm{min}$, $\gamma_\mathrm{max}$, $t_\mathrm{var}$, and $\delta$ are initially kept constant to eliminate degeneracies, the initial values were selected from previous SSC modelling of multi-wavelength observations of Mkn~501 \citep{1998ApJ...509..608T} and are not directly affecting the shape of the predicted average TeV spectrum.

\setlength{\tabcolsep}{1.5pt}
\renewcommand{\arraystretch}{1.5}
\begin{table}
\caption{List of parameters and values that enter the SSC model calculations. For the parameters left free in the fit, the resulting $1~\sigma$ uncertainties are listed as well. See Sect. \ref{sec:TeV-SSC} for further details.}
\label{tab:SSC-pars}
\begin{tabular}{lll}
\hline
Parameter & Value  & Description \\ \hline
$K$ & (4.8 $\pm$ 1.2)$\times 10^3$ cm$^{-3}$ & Normalisation \\ 
$\gamma_{min}$ & 3$\times 10^3$ & Minimum Lorentz factor \\
$\gamma_{max}$ & 10$^9$ & Maximum Lorentz factor \\ 
$\gamma_{c}$ & (3.7 $\pm$ 0.6)$\times 10^6$ & Cut-off Lorentz factor \\
P & -2.56 $\pm$ 0.03 & Power-Law index \\
t$_{\text{var}}$ & 12 days (0.14 pc) & Variability time (radius) \\
$\delta$ & 15 & Doppler factor \\
$B_\mathrm{SSC}$ & (3 $\pm$ 0.3) mG & Magnetic field \\
z & 0.034 & Redshift of the source \\ \hline
\end{tabular}
\end{table}

The best-fit SSC model is shown in Fig.~\ref{fig:SSC-fit-501}. The model curve represents the data quite well as the 
$\chi^2(df)=32.5(31)$ indicates. The two broad peaks in the SED are reproduced. The TeV flux predicted by the model is slightly larger than the archival data from 2009. However, this is expected given the relatively low-flux state of the source during the TeV observations in 2009.

\section{Cascade simulations and halo templates for Mkn~501}
\subsection{ELMAG simulations of cascade emission}
\label{sec:ELMAG}
The secondary gamma-ray emission is linked to the primary emission in the absorbed part of the injection spectrum determined in the previous section \ref{sec:TeV-SSC}. The optical depth is estimated using the model by \citet{2010ApJ...712..238F}.
The detailed three-dimensional treatment of pair production and inverse Compton scattering in a magnetised medium is implemented in the Monte Carlo-based ELMAG (v3.03) program \citep{2020CoPhC.25207163B}.

The RMS of the intervening magnetic field $B_\mathrm{rms}$ is modelled as the superposition of n$_{\mathrm{k}}$ transverse Fourier modes $B_j$ for $k_\mathrm{min}<k_j<k_\mathrm{max}$ that follow a Kolmogorov-type spectrum with index $\gamma=5/3$  and a vanishing helicity:
$$B_\mathrm{j} = B_\mathrm{{min}} \left( \frac{k_j}{k_{min}}  \right) ^{-\dfrac{\gamma}{2}}.$$
Here, B$_{\mathrm{min}}$ denotes the amplitude of the lowest Fourier mode, which is determined by normalising the chosen value of B$_{\mathrm{rms}}$ 
$$B_{\mathrm{min}} = \frac{B_{\mathrm{rms}}}{\sqrt{\sum_{j=1}^{n_k}\left(\frac{k_j}{k_{\mathrm{min}}}\right)^{-\gamma}}}$$
The minimum and maximum wavenumbers 
k$_{\mathrm{min}}$ and k$_{\mathrm{max}}$ are related to the  coherence length parameter 
$\ell_{\mathrm{C}}$ as:
$$k_{\mathrm{min}} = \frac{2\pi}{5 \ \ell_C} \hspace{50pt} k_{\mathrm{max}} = \frac{2\pi}{5\times10^{-4} \ \ell_C}$$
For further details on the simulations and the modelling of the turbulent magnetic field, we refer to the original publication \citep{2020CoPhC.25207163B}.

The secondary particles with an energy larger than
$10^8$~eV are tracked until they cross a sphere centred on the source and with a radius given by the luminosity distance to the observer.
Their arrival direction, energy and time delay are registered and stored together with statistical weights to take into account the injection energy spectrum and a conical jet geometry with a half opening angle $\theta_{\mathrm{jet}}$, inclined with 
respect to the line of sight by the angle $\theta_\mathrm{obs}$.
\subsection{Simulated halo templates from Mkn~501}
\label{sec:templates}
The simulation as described above is run for a redshift $z=0.034$ on a logarithmic grid of 24$\times$9 values of $B_\mathrm{rms}$ and $\ell_C$, covering the range from $\SIrange{2.5e-18}{e-13}{G}$ and
$\SIrange{e-3}{10}{\mega\parsec}$ respectively, with $\theta_\mathrm{jet}=3^\circ$ and $\theta_\mathrm{obs}=0^\circ$. 

For each combination of $B_\mathrm{rms}$ and $\ell_C$, 
up to $2\times 10^6$ photons are injected between $0.3~\mathrm{TeV}$ and $100~\mathrm{TeV}$. The injected photon spectrum is parametrised to follow the SSC model developed in Sect.~\ref{sec:TeV-SSC} using a power-law with a modified exponential cut-off:

\begin{ceqn}
\begin{equation}
    \label{eq:inj_power}
    \frac{dN}{dE} = N_0 \left(\frac{E}{E_0}\right)^{-\Gamma}~\exp{\left( -\frac{E}{E_c} \right)^{\beta}},
\end{equation}
\end{ceqn}
with $N_0=1.04 \times 10^{-5}~\mathrm{(TeV~s~cm^2)^{-1}}$, $E_0 = 1~\mathrm{GeV}$, $\Gamma=1.72$, $E_c=0.33~\mathrm{TeV}$, and $\beta=0.37$.

Here, $E_c$ is the characteristic scale of the sub-exponential turnover rather than a hard upper cut-off; with $\beta=0.37$, the adopted spectrum retains a multi-TeV tail. After EBL absorption is included, the pair-injection power, given by the integrand of Eqn.~\ref{eqn:liso}, peaks at $\approx1.5$~TeV, comparable to the $\sim2$~TeV inverse-Compton peak measured during the extreme 2012 state of Mkn~501, although emission beyond 10~TeV has been observed during exceptional flares \citep{2018A&A...620A.181A,1999A&A...349...11A}. Such hard states can enhance the highest-energy part of the pair injection and shift part of the cascade production closer to the source; because they are short-lived and the SSC model is derived from a long-term average over contemporaneous hard-X-ray and gamma-ray observations, their contribution is diluted, although residual spectral-shape variability remains an additional systematic uncertainty.

\begin{figure}
\centering
\includegraphics[trim=0 0 0 10,clip, scale=0.45]{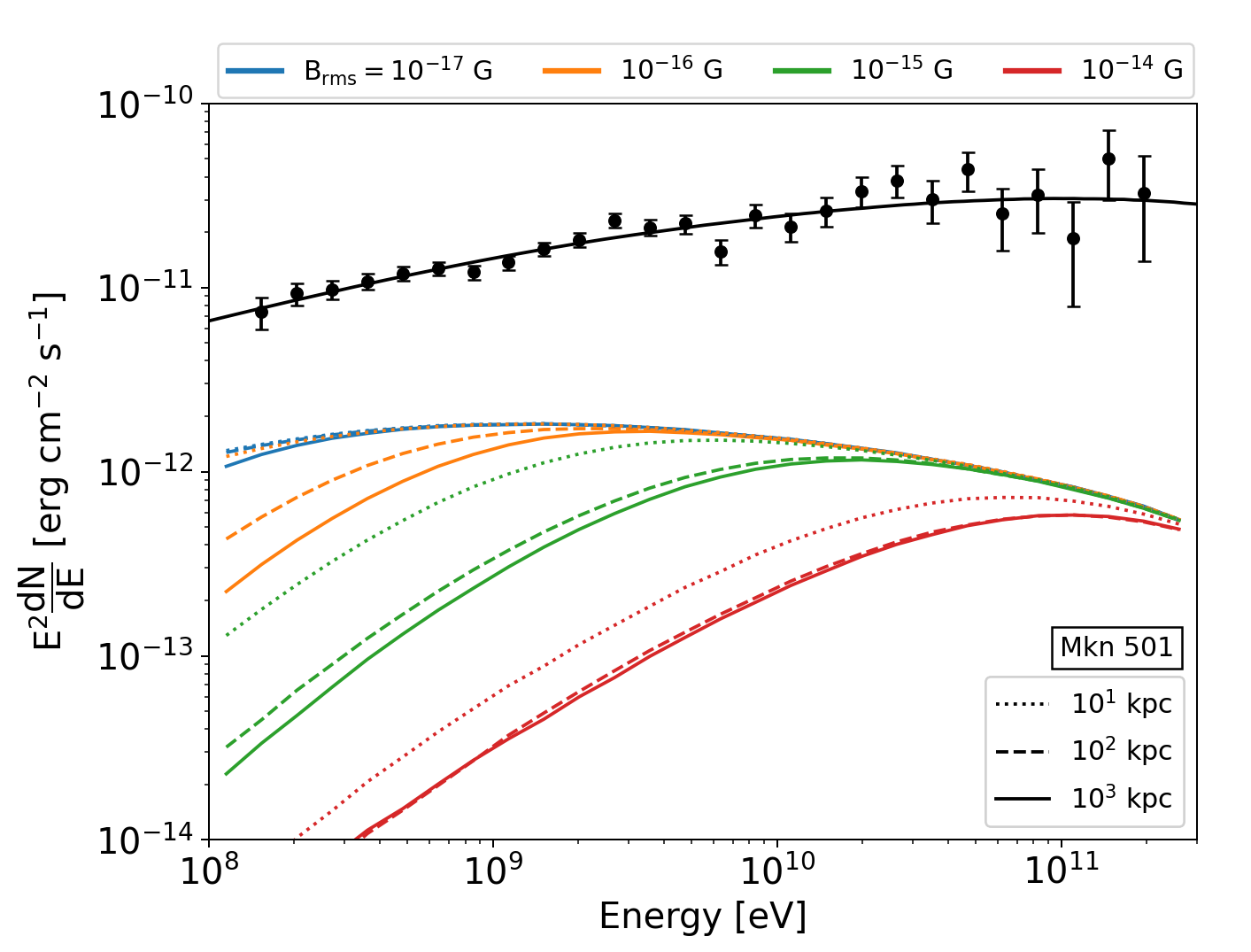}
\caption{Spectral energy distribution of the simulated halo templates for Mkn~501 at different $B_{\mathrm{rms}}$ and $\ell_C$ values. Black dots show the SED from $\it{Fermi}$-LAT and the black line represents the SSC fit which has been used to estimate the energy injected into the simulations. Here it is assumed that no additional energy loss mechanisms are present in the production of the secondary emission.}
\label{fig:mkn501_sed}
\end{figure}

The secondary photons registered with a time-delay of up to $10^7$ years are binned in energy and angular coordinates centred on the ROI as defined for the \textit{Fermi}-LAT analysis (see Sect.~\ref{sec:gamma-obs}). 
The treatment of a finite source duty-cycle is described in Appendix \ref{app:duty-cycle}, and the time-delay simulations are presented in Appendix \ref{app:time-delay}.
As required by the Fermi-analysis tools, the sum of all weights in the resulting halo template cube is normalised to unity. 
The resulting SED of the simulated halo is shown in Figure \ref{fig:mkn501_sed} for selected values of $B_\mathrm{rms}$ (colours) and $\ell_C$. Note, the secondary emission is normalised to the injection spectrum shown as a solid line. The results demonstrate that the secondary emission is suppressed with increasing $B_\mathrm{rms}$. The effect is very pronounced for $B_\mathrm{rms}\gtrsim 10^{-16}~\mathrm{G}$, where the secondary photons with energies $\lesssim 1~\mathrm{GeV}$ are deflected too much to be registered within the ROI of $10^\circ \times 10^\circ$.
The effect of increasing the coherence length $\ell_C$ similarly leads to a suppression of the secondary flux at lower energies.

The angular extension of the emission is conveniently characterised by the containment radius $r_{68}$ which is obtained by summing the weighted flux for a fixed energy interval over discs with increasing radius until the radius is found where $68~\%$ of the total flux is contained.

\begin{figure}
\centering
\includegraphics[trim=0 0 0 10,clip, scale=0.48]{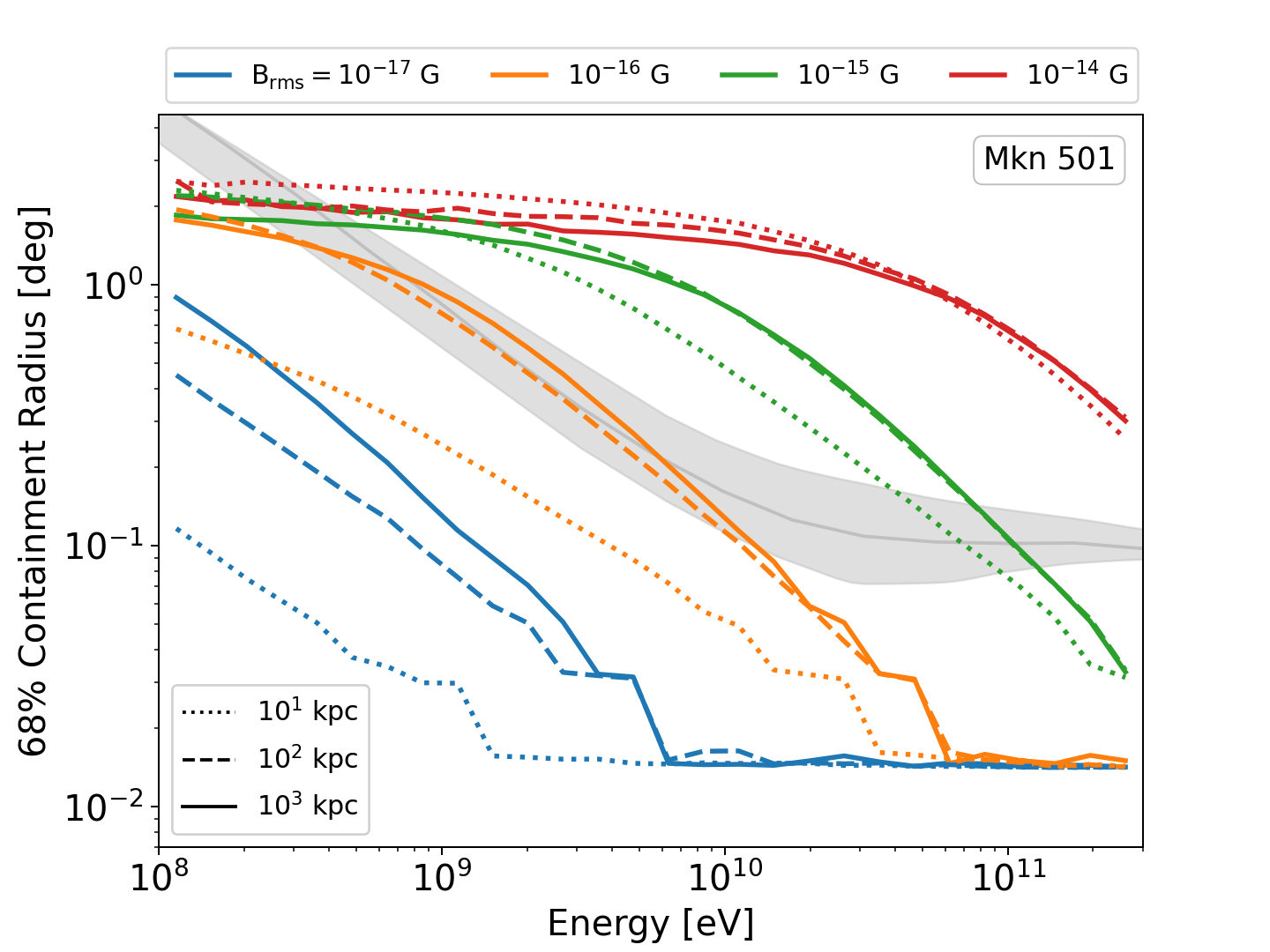}
\caption{Containment radius r$_{68}$ in degrees where 68 percent of the total cascade flux is contained, calculated for the templates simulated for Mkn 501 using different $B_{\mathrm{rms}}$ strengths and coherence length values as labelled. The gray band represents the $\it{Fermi}$-LAT 68~\% PSF, for $\texttt{FRONT}$ and $\texttt{BACK}$ conversion from the latest IRF \texttt{P8R3$\_$SOURCE$\_$V3}.}
\label{fig:mkn501_68}
\end{figure}

The resulting estimates of $r_{68}$ for selected values of $B_\mathrm{rms}$ and $\ell_C$ are shown in Fig.~\ref{fig:mkn501_68}.
As expected, the angular extension increases with increasing magnetic field. The maximum extension saturates for low energies at $\approx 2.5^\circ$ where the effect of the ROI boundaries is limiting the summation.
At high energies, the minimum containment radius that can be reliably estimated from the halo template is limited by the bin size of $0.02^\circ$. 
When comparing the apparent halo size for different values of $\ell_C$ and fixed $B_\mathrm{rms}\lesssim \SI{e-15}{G}$, the extension increases with increasing $\ell_C$.
At larger $B_\mathrm{rms}$, the differences vanish or are even inverted. 
This is most likely related to the limited size of the ROI. 

\section{Searching and characterising cascade emission around Mkn~501}

\subsection{Hypotheses-testing with the likelihood-ratio method}
\label{sec:LRT}

The likelihood $L$ of the binned data counts $n_i$ for a given model expectation $m_i$ in bin $i$ is calculated under the assumption of Poissonian count process \citep{1996ApJ...461..396M}:
$$ L = \prod_{i} \frac{m_{i}^{n_{i}} e^{-m_{i}}}{n_{i}!}. $$
The underlying model parameters are estimated by maximizing $\mathcal{L}=\ln L$
\begin{ceqn}
\begin{equation}
\label{eq:Like}
    \mathcal{L} = \sum_{i} n_{i} \ln(m_{i}) - N_\mathrm{pred}, 
\end{equation}
\end{ceqn}
with the total number of photons predicted $N_\mathrm{pred}=\sum_i m_i$. Note, the factorial $n!$ is dropped since it does not depend on the parameters of the model. 

Here, two different, nested models are used which follow from the two hypotheses: 
\begin{description}
    \item[\bf Null hypothesis ($H_0$):] The gamma-ray emission in the ROI is produced by the known sources of the 4FGL catalogue, the isotropic background and galactic foreground emission. 
    \item[\bf Alternative hypothesis ($H_1$):] In addition to the known sources of gamma-ray emission, an extended halo source centred on Mkn~501 is expected.
\end{description}
The resulting models are fit to the data by maximizing the log-likelihoods $\mathcal{L}_0$ and $\mathcal{L}_1$ for the two hypotheses respectively \footnote{The products required to replicate the analysis are available at \url{https://github.com/matesw/IGMF_repo}}. We introduce a test statistic based on the likelihood ratio to estimate the statistical preference of the alternative hypothesis against the null hypothesis:
\begin{ceqn}
\begin{equation}
\label{eq:TS}
TS(B,\ell_C) = -2 \left(\mathcal{L}_{0} -\mathcal{L}_1(B,\ell_C)\right)
\end{equation}
\end{ceqn}
The value of $TS$, assuming that the models are nested and the additional parameter (normalisation of the halo emission $\varepsilon_\mathrm{eff}$) is not at the boundary of the parameter interval, is expected to follow a $\chi^2(df)$ probability density function with $df=1$ degrees of freedom \citep{Wilks:1938dza}.

\subsection{Null hypothesis H$_0$}
\label{sec:model-base}
The model for the null hypothesis includes all catalogued sources within a sky region of 
$15^\circ\times 15^\circ$ 
centred on Mkn~501. The sky positions for these sources are kept fixed while the spectral model parameters are left to vary for all sources with an offset angle to Mkn~501 less than $5^\circ$ or with a $TS\geq25$. In addition to the point sources, the model includes the isotropic background and the Galactic foreground emission with normalisations left free in the fit procedure. 

In Table \ref{tab:L1-model}, the fit results for the eight brightest sources and the diffuse components are listed for the best-fitting model with $\mathcal{L}_0=\num{-1027889.69}$. 

\subsection{Alternative hypothesis H$_1$}
\label{sec:H1}
The model used for $H_1$ includes all sources of the model for $H_0$ and a template cube from the simulations (see Sec.~\ref{sec:ELMAG}) with a free normalisation $\varepsilon_\mathrm{eff}>0$. 
The efficiency is chosen such that for $\varepsilon_\mathrm{eff}=1$, the secondary flux corresponds to the simulated cascade emission for the given injection spectrum. 

For the alternative hypothesis, the model includes two additional free parameters ($B_\mathrm{rms}$ and $\ell_C$) that are not present for the null hypothesis. The library of templates  
generated with ELMAG simulations is used as input for the fits and the resulting $TS$ is calculated for each simulated combination of $B_\mathrm{rms}$ and $\ell_C$. 

In Table~\ref{tab:L1-model}, the resulting values for $N_\mathrm{pred}$ are listed, including the best-fitting value for the halo source (1402 photons) with $\mathcal{L}_1=\num{-1027870.99}$ which translates into a $TS=\mathbf{37.4}$ for the likelihood ratio test. 

\begin{table}[htbp]
\small
\caption{ROI model for the alternate hypothesis, including the extended template component. The 4FGL catalogue name is given. The global log-likelihood value obtained is $\mathcal{L}_1 = \num{-1027870.99}$}
\label{tab:L1-model}
\centering
\begin{tabular}{lllll}
\hline
4FGL name & Assoc. name & Offset ($^\circ$) & $H_0$: $N_\mathrm{pred}$ & $H_1$: $N_\mathrm{pred}$ \\
\hline
J1653.8+3945 & Mkn 501 (pt)        & 0.00 & 21541 & 21147 \\
J1642.9+3948 & 3C 345              & 2.10 & 18983 & 18804 \\
J1640.4+3945 & NRAO 512            & 2.57 & 12911 & 13031 \\
J1648.2+4232 & IVS B1646+426       & 2.98 & 3803  & 3819 \\
J1635.2+3808 & 4C +38.41           & 3.96 & 64544 & 64609 \\
J1709.7+4318 & B3 1708+433         & 4.62 & 8340 &  8345 \\
J1635.6+3628 & MG3 J163554+3629    & 4.85 & 2507 &  2529 \\
J1727.4+4530 & S4 1726+45          & 8.42 & 2378 &  2372 \\
J1653.8+3945 & Mkn 501 (ext)       & ---  & ---  &  1402 \\
isodiff      &  ---                & ---  & 109037 & 107048 \\
galdiff      &  ---                & ---  & 108093 & 108243 \\
\hline
\end{tabular}
\end{table}

\begin{figure}
   \centering
   \includegraphics[trim=0 0 0 0,clip,scale=0.48]{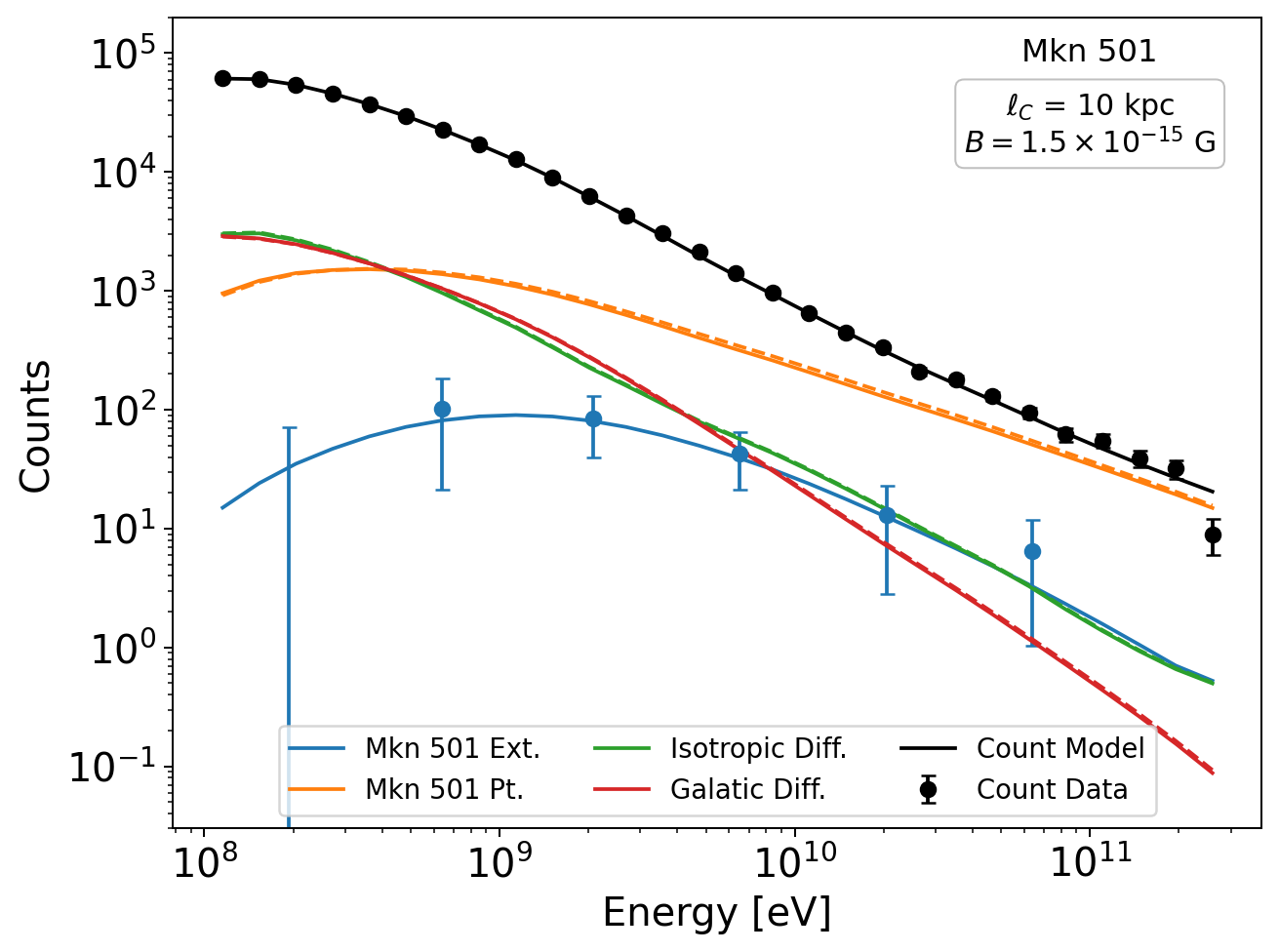}
   \caption{Count spectra for the ROI of Mkn~501, including the contribution from the most significant components of the model. \textit{Mkn~501 Pt.} and \textit{Mkn~501 Ext.} represent the point and extended sources respectively. 
   Count spectra are retrieved for the best-fit combination of B$_{\mathrm{rms}}$ and $\ell_C$ parameters. Dashed lines show the count spectra from the analysis performed for the $H_0$ model, showing that the counts which are attributed to the extended templates come primarily from the point source and galactic diffuse components.}
   \label{fig:mkn501-count-spec}
\end{figure}

In Fig. \ref{fig:mkn501-count-spec}, the photon count spectrum is shown for the central $4^\circ \times 4^\circ$ region of the ROI. The solid black line is the sum of all model components included in the alternative hypothesis $H_1$. The point-sources in the model and the diffuse and isotropic backgrounds are subtracted to estimate the number of photons that are assigned to the extended halo of Mkn~501. The resulting excess spectrum is rebinned  to obtain the blue data points. The energy count spectrum of the halo component shows a harder spectrum than the other components with the main signal accumulated between approximately $0.5$~GeV and $30$~GeV.

\begin{figure}
   \centering
   \includegraphics[trim=5 0 0 0,clip,scale=0.48]{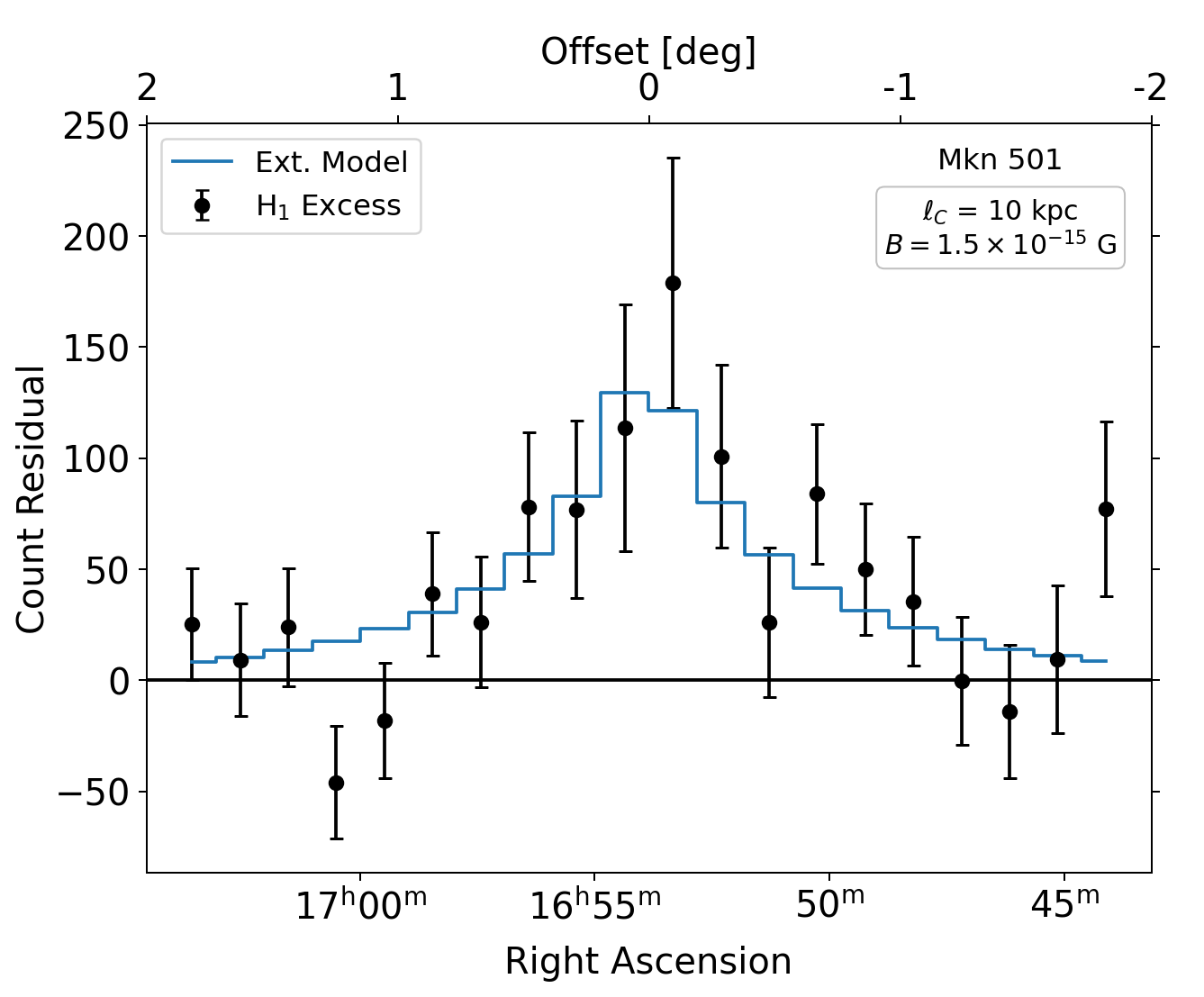}
   \caption{Spatial count residuals obtained in projection along the right ascension for $4^\circ \times 4^\circ$ section centred in the direction of blazar Mkn~501. The energy range is between 0.5 and 30 GeV. The residuals are calculated by subtracting from the binned data the count maps generated for all sources except for the extended source in the $H_1$ model. The blue histogram is the best-fitting model for the extended source around Mkn~501.}
   \label{fig:mkn501-count-res}
\end{figure}

In Fig. \ref{fig:mkn501-count-res}, the spatial model for the hypothesis $H_1$  (without the extended source model) is subtracted from the data cube in the energy range from $0.5$~GeV to $30$~GeV.
The resulting excess counts projected in bins of right ascension are shown in black ($H_1$ model), overlaid with the best-fitting model as a blue histogram. The excess counts estimated this way are in agreement with the expected shape of the extended emission with a FWHM of $\approx 1^\circ$ for the best-fitting model.

\subsection{Null hypothesis simulations}
As described above (Sect.~\ref{sec:LRT}), the Wilks' theorem can be applied under certain conditions. In the case considered here, the additional parameter $\varepsilon_\mathrm{eff}\ge 0$ is at the boundary of the parameter space. In this case, dedicated simulations under the null hypothesis are needed to sample and characterise the probability density function of the test statistics (likelihood ratio, see Eqn.~\ref{eq:TS}), see \citet{Protassov_2002} for further discussions.  

\begin{figure}
   \centering
   \includegraphics[trim=0 0 0 0,clip,scale=0.58]{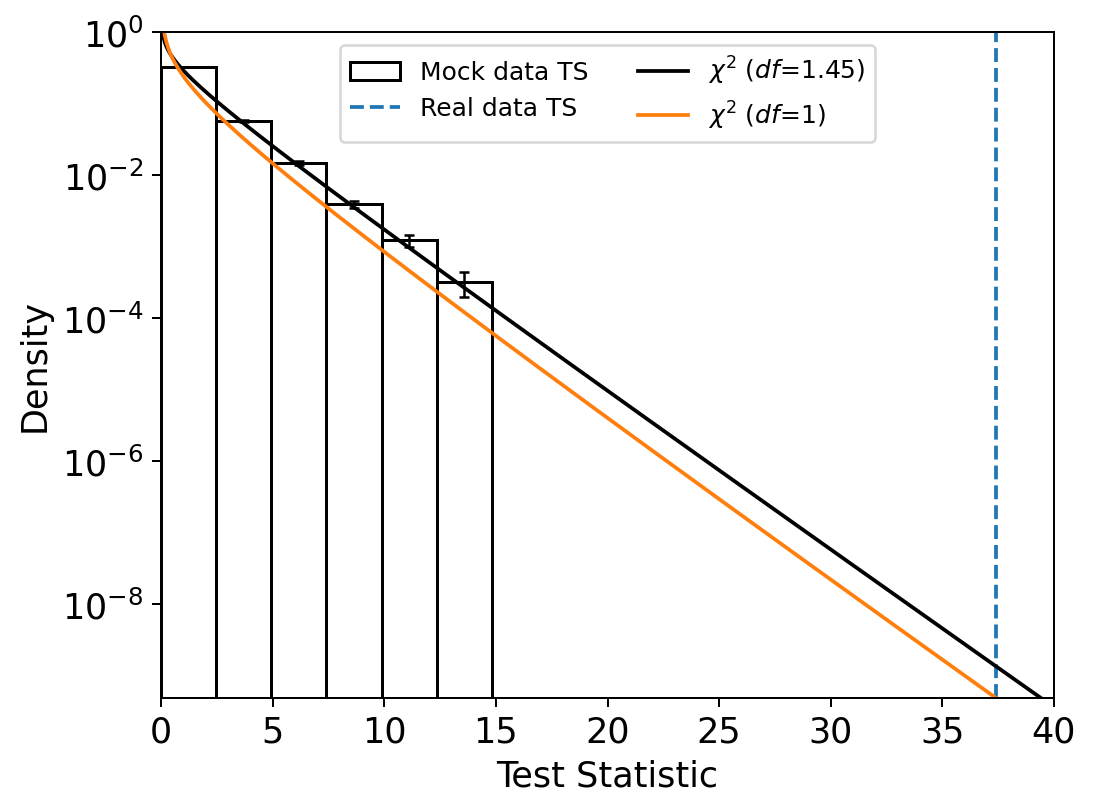}
   \caption{Distribution of test statistics (TS) values found from analysing \num{10000} mock datasets simulated under the $H_0$ hypothesis. 
   Overlaid are $\chi^2$-functions for $df=1$ and $df=1.45$. The dashed line shows the TS obtained from our fit to the actual data.}
   \label{fig:mockTS-501}
\end{figure}

These simulations are carried out by re-sampling the best-fitting $H_0$ model assuming Poisson count statistics for the binned data cube. In total, \num{10000} mock data sets are generated and analysed in the same way as the data set described above. The model for the $H_1$ hypothesis is chosen to have 
$B_\mathrm{rms}=\SI{1.5e-15}{G}$ and 
$\ell_C=\SI{10}{\kilo\parsec}$. The analysis of the simulated data sets is limited to this best-fit IGMF configuration due to the large computational time required. As a cross-check, we have confirmed that for one more combination of $B_\mathrm{rms}=1.5$ fG and $\ell_\mathrm{C}=1$ Mpc, we find a value of $df=1.48$, similar to the one used here.
The resulting normalised TS distribution from the simulations is presented in Fig.~\ref{fig:mockTS-501}. The relative frequency distribution follows closely a $\chi^2$-function with one degree of freedom as predicted by the Wilks' theorem. However, there are some deviations visible and we fit a more generalized $\chi^2$-function to the data with the degrees of freedom as a free parameter. The best fit with $df=1.45$ is in very good agreement with the simulations and is used to extrapolate towards larger TS values as found in the data ($TS_\mathrm{max}=$37.4, marked as a blue dashed vertical line). The corresponding significance for finding such a large value of $TS$ is given by the survival probability of the $\chi^2$-function with $df=1.45$ to be $P(\chi^2(df=1.45)>TS_\mathrm{max})=\num{2.6e-9}$, which translates into $5.8~\sigma$ deviation from the null hypothesis.

\subsection{Hypothesis testing with additional free parameters}
The model for hypothesis $H_1$ includes additional free parameters $B_\mathrm{rms}$, $\ell_C$ that are not present in the model for $H_0$\footnote{The 
	parameters related to the jet geometry and the spectral shape of the magnetic field fluctuations had been fixed \textit{a priori} and were not
varied during the likelihood maximisation}. In this case, the survival probability $P(TS(B,\ell_C)>TS)$
needs to be corrected for the additional number of trials introduced \citep{4f1875a3-4f5f-31ca-b6dd-ea6711efe2a2}. 

We performed our analysis scanning 24 bins for B$_{\mathrm{rms}}$ and 9 bins for $\ell_C$, therefore, we conducted 216 searches for each source. The resulting plane of $TS$-values is shown in Fig.~\ref{fig:mkn501-TS-grid} with the peak value $TS_\mathrm{max}=$37.4 marked by a black cross.  

Following the approach suggested by \citet{2010EPJC...70..525G} the resulting survival probability is corrected using the effective number of trials $\mathcal{N}$
$$P(TS(B,\ell_C) > TS_{max}) \approx P(\chi^2_{df} > TS_{max}) + \mathcal{N} P(\chi^2_{df+1} > TS_\mathrm{max}).$$

We estimate $\mathcal{N}\approx 12$ by counting the 
number of search bins with a $TS$-value larger than $TS_\mathrm{max}-2.3$, as indicative for a $68~\%$ uncertainty region. 
With these numbers, the corrected probability is $P(TS>TS_\mathrm{max}) = \num{2.5e-7}$, which translates into a corrected significance of $5.0~\sigma$.
\footnote{When correcting the survival probability using the number of trials $n\approx 74$ estimated from eqn. 9 in \citet{2010EPJC...70..525G} in conjunction with the simple binomial correction $1-(1-P(\chi^2(df=1.45)>TS_\mathrm{max}))^n$, a very similar result is obtained.}

\section{Results}
\label{sec:results}
\begin{figure}
   \centering
   \includegraphics[trim=0 0 0 0,clip,scale=0.55]{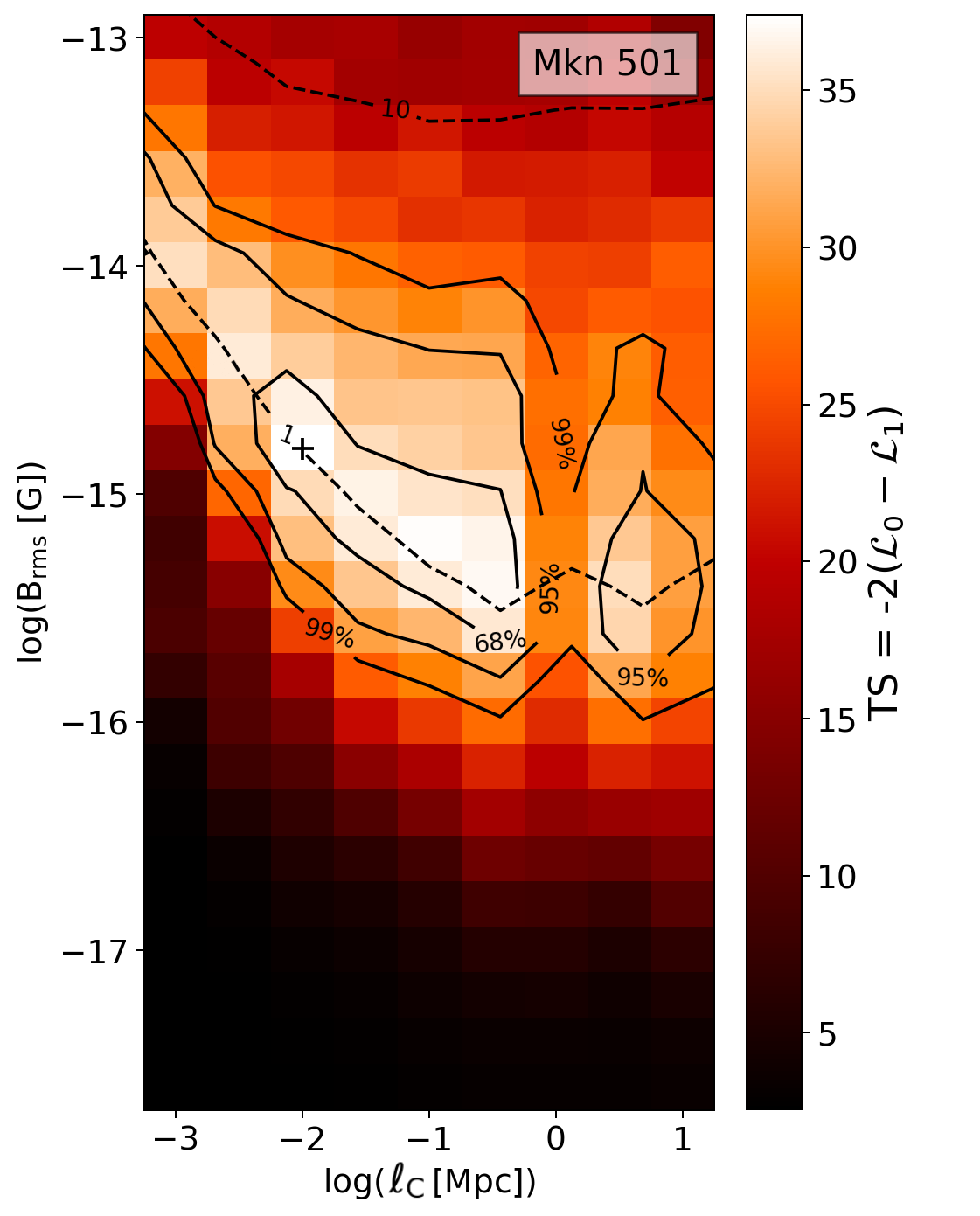}
   \caption{Test Statistic obtained from our fit to the Mkn501 gamma-ray data, using the different $B_{\mathrm{rms}}$ strengths and coherence lengths $\ell_{\mathrm{C}}$. The plus symbol marks the parameter combination that resulted in the highest TS value with TS$_{\mathrm{max}}$ = 37.4. Dashed lines show the contours where the calculated efficiency $\varepsilon_{\mathrm{eff}}$ is equal to 1 and 10 as labelled.}
   \label{fig:mkn501-TS-grid}
\end{figure}

The TS values obtained from fitting our models to the Mkn 501 $\it{Fermi}$-LAT data are presented as a two-dimensional map in Fig.~\ref{fig:mkn501-TS-grid}.
We find a set of parameters where the Test Statistic is maximized with TS$_{\mathrm{max}}$ = 37.4, for B$_{\mathrm{rms}}$=1.5$\times$10$^{-15}$ G and $\ell_\mathrm{C}$ = 10 kpc, with an efficiency $\varepsilon_\mathrm{eff} = 0.97 \pm 0.11$. 
These results are obtained for the templates that consider TeV absorption from the Gilmore EBL model, and using a spatial binning of 0.02 degrees. 

We include black solid-line contours that define the 68, 95 and 99 coverage probabilities, as calculated for a joint estimation of two parameters \citep{ParticleDataGroup:2024cfk}.
The shape of the contours indicates an anti-correlation between coherence length and magnetic field strength. This is expected as the angular size of the secondary emission
scales with $\ell_C^{1/2}~B_\mathrm{rms}$.  We extract the covariance matrix from the two-dimensional TS map and find 
$\log_{10}(B_\mathrm{rms}/\mathrm{fG})=0.18\pm0.46$ ($B_\mathrm{rms}=1.5^{+2.9}_{-1}~\mathrm{fG}$),
$\log_{10}(\ell_\mathrm{C}/~\mathrm{kpc}) = 1\pm 0.9$ ($\ell_\mathrm{C}= 10^{+64}_{-9}~\mathrm{kpc}$) with a correlation coefficient of $\rho=-0.72$.  Note however, the 
bounding values of the asymmetric 68~\% contour include a larger range of values than the one estimated from the covariance matrix. We therefore quote the 
corner values of the contour as uncertainty: $B_\mathrm{rms}=1.5^{+2.0}_{-1.3}~\mathrm{fG}$, $\ell_\mathrm{C}=10^{+490}_{-5}~\mathrm{kpc}$.
Dashed-line contours show the points where the efficiency $\varepsilon_\mathrm{eff}$ calculated is equal to 1 and 10 as labelled.

\subsection{SED of the cascade emission}
\begin{figure}
   \centering
   \includegraphics[trim=0 0 0 0,clip,scale=0.46]{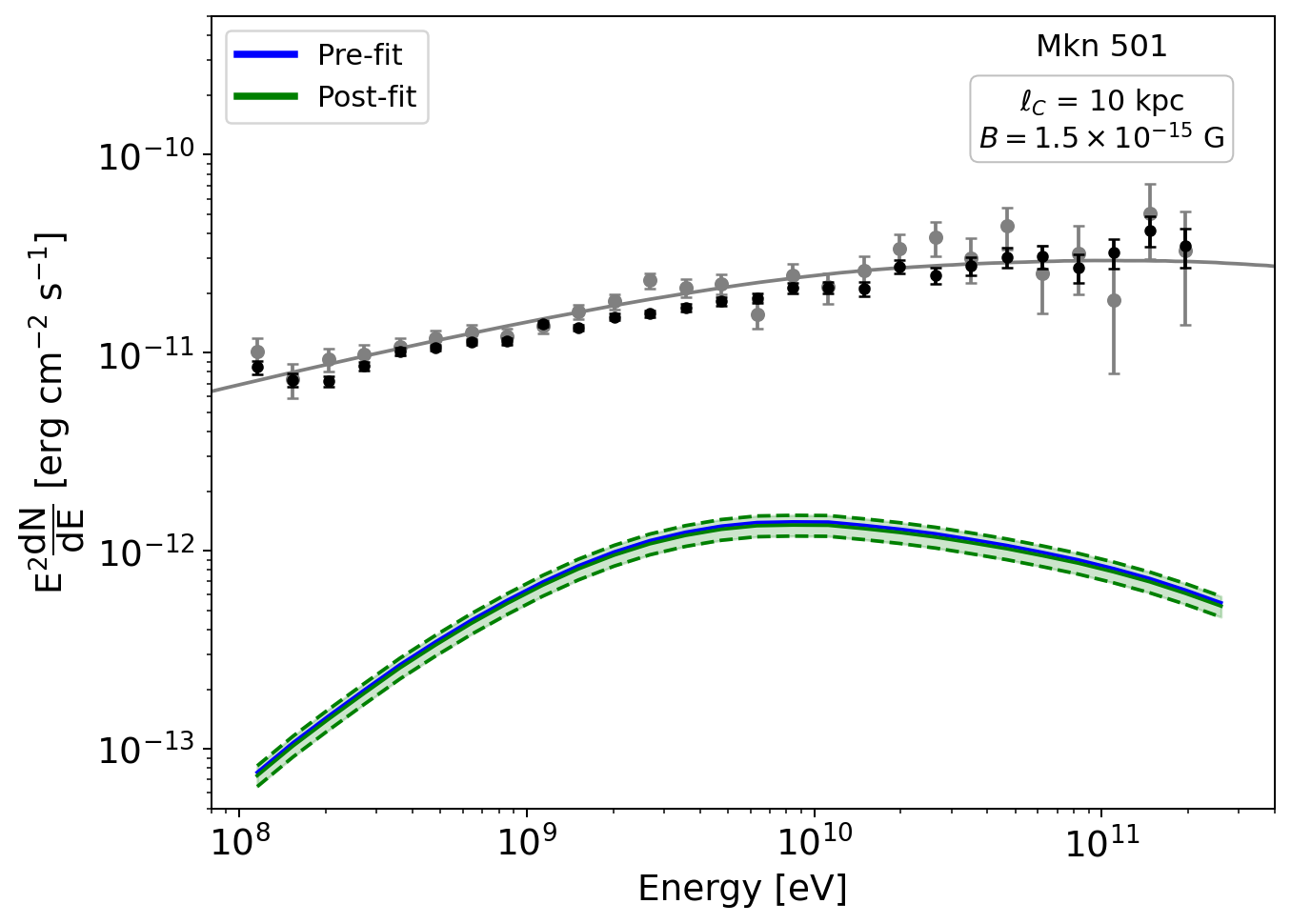}
   \caption{SED of Mkn 501 including the expected SED for the best-fit template (blue) and the one normalised after fitting (green, see text). Black data points represent the full \textit{Fermi}-LAT 14 year data sample, and in dark gray we show the contemporaneous dataset which was used to fit the SSC model. }
   \label{fig:SED-fitted-501}
\end{figure}

The best-fitting template for $B_\mathrm{rms}=\SI{1.5e-15}{G}$ and $\ell_C=\SI{10}{\kilo\parsec}$ is shown as a blue line in Fig~\ref{fig:SED-fitted-501} together with the SED obtained from the contemporaneous data-set as grey data points (see Sect.~\ref{sec:contemporaneous-obs}). In addition, the black data points are the result of reconstructing the SED using the entire data-set covering 14 years of observation with \textit{Fermi}-LAT. 

The best-fitting value for $\varepsilon_\mathrm{eff}=$0.97 $\pm$ 0.11 which is shown as the \textit{post-fit} (green) band in Fig.~\ref{fig:SED-fitted-501}. As can be seen from the figure, the SED of the cascade emission peaks at $\approx \SI{8}{\giga\eV}$ where it contributes $\approx 10~\%$ of the flux assigned to the point source. 

\subsection{Systematic uncertainties}
\label{sec:systematics}
In the following, we are considering the systematic uncertainties on the reconstructed $B_\mathrm{rms}$ and efficiency $\varepsilon_\mathrm{eff}$ resulting from the systematic uncertainties of the instrument and uncertainties related to the underlying assumptions of the source and the EBL. 
\paragraph{Instrumental uncertainties:}
The TS reported for the hypothesis test is obtained with the entire data set which includes different event types with widely different point spread functions (PSF). 
Each event is classified according to the location of pair conversion (FRONT and BACK) and additionally, for each event a PSF type is assigned (PSF0-PSF3) such that events 
of the PSF3-type have the best angular resolution. In order to test the robustness of the signal, we have carried out the analysis for FRONT, BACK, PSF0-PSF3 events separately and
report in Table~\ref{tab:front-back} the resulting log-likelihood and TS-values. About 55~\% of the events are FRONT-converted while the TS is slightly larger for the BACK-converted events
which have on average a degraded PSF. The PSF0-3 event types split the data into roughly equal shares. The TS values for the different PSF types are close to each other with the 
exception of PSF1. Overall, the excess is seen across different event types in a consistent way such that we can rule out a singular contribution to the overall excess.

Systematic uncertainties in the observational data arise primarily from the Fermi-LAT calibration and are propagated through the instrument response functions (IRFs). We assess uncertainties associated with the effective area $A_\mathrm{eff}$, and PSF using the bracketing method \citet{2012ApJS..203....4A}. For $A_\mathrm{eff}$, we use the bracketing IRFs provided by the Fermi-LAT collaboration, which contain appropriately scaled effective areas. \footnote{\url{https://fermi.gsfc.nasa.gov/ssc/data/analysis/scitools/Aeff_Systematics.html}} For the PSF, the \texttt{SCORE} and \texttt{STAIL} parameters defining the PSF model \footnote{\url{https://fermi.gsfc.nasa.gov/ssc/data/analysis/documentation/Cicerone/Cicerone_LAT_IRFs/IRF_PSF.html}} are scaled by factors of 0.95 and 1.05 to produce narrower and wider PSFs, respectively. We repeat the analysis with each modified IRF, fixing $\ell_\mathrm{C}$ to 10 kpc.

The $A_\mathrm{eff}$ bracketing has a negligible effect on the results (Table~\ref{tab:systematics}). In contrast, the PSF variations substantially affect the inferred values of $B_\mathrm{rms}$ and $\varepsilon_\mathrm{eff}$. Re-optimizing both parameters for each bracketing IRF, the narrow PSF yields $\varepsilon_\mathrm{eff}=5.51$ and $\log_{10}(B_\mathrm{rms}/\mathrm{G})\simeq-15.4$. The wide PSF instead results in $\varepsilon_\mathrm{eff}=0.19$ and $\log_{10}(B_\mathrm{rms}/\mathrm{G})\simeq-14.0$. We therefore report PSF-related systematic uncertainties of $\varepsilon_\mathrm{eff}
 = 0.97^{+4.54}_{-0.78}$ and $\log_{10}(B_\mathrm{rms}/\mathrm{G})
 = -14.8^{+0.8}_{-0.6}$.

 For both bracketing IRFs, the maximum log-likelihoods under $H_0$ and $H_1$ are lower than for the nominal IRF, indicating a poorer overall description of the ROI. Nevertheless, the relative preference for the cascade component increases from $TS_\mathrm{nominal}=37.4$ to $TS_\mathrm{narrow}=74.5$ for the narrow PSF and decreases to $TS_\mathrm{wide}=21.9$ for the wide PSF. This behaviour reflects the redistribution of emission between the point source and cascade components. A wider PSF attributes more emission in the tails to Mkn~501, whereas a narrower PSF leaves a larger residual to be modelled by the cascade template. For a more detailed revision of the PSF bracketing, see Appendix~\ref{app:PSF-bracketing}.

\paragraph{Theoretical uncertainties:}
The absorption of TeV emission is calculated for an assumed EBL model. The models differ in the predicted optical depth. This in turn leads to variations 
of the injected power in electron/positron pairs which leads to different reconstructed efficiency values $\varepsilon_\mathrm{eff}$ while $B_\mathrm{rms}$ remains unchanged. 
In Table~\ref{tab:EBL_eff}, we report the resulting
best-fit values for $\varepsilon_\mathrm{eff}$ when using the six different EBL models which are implemented in ELMAG. 

The observed TeV flux is known to vary on all observed time-scales with observational indication of a decorrelation at time-scales larger than 
$\mathcal{O}(10~\mathrm{yrs})$ \citep{2011A&A...531A.123K}. We estimate the variation of the reconstructed efficiency from the light curve of the TeV flux observed with 
the FACT telescope \citep{2021A&A...655A..93A}. The resulting efficiency varies by $40~\%$ when considering the central 68 percentile flux interval of the FACT light curve. The
reconstructed magnetic field remains unchanged. In addition to flux changes, Mkn~501 is known to undergo spectral changes during high states as well, first observed during the 1997 flare 
\citep{1999A&A...349...11A}. The relative duty cycle of these exceptional high-flux states is $O(10^{-2})$ and the resulting effect on the reconstructed parameters is negligible. 

The source is assumed to inject TeV photons during its duty cycle $t_\mathrm{dc}$. As shown in Appendix \ref{app:time-delay}, the required duty-cycle $t_\mathrm{dc}\gtrsim 4.5\times 10^{4}$ years. When 
varying the duty cycle between the lower bound and $10^7$~years, the resulting efficiency $\varepsilon_\mathrm{eff}$ is lowered to $0.8$ and $B_\mathrm{rms}$ is reduced by a factor of 25. 

Finally, the choice of the jet opening angle and the alignment of the jet with the line of sight is only weakly constrained. The opening angle scales with the bulk Lorentz factor $\Gamma$ for small jet angles $\theta_\mathrm{jet}\propto \Gamma^{-1}$. The jet alignment angle $\vartheta$ is fixed to $0^\circ$ and the opening angle is varied between $1^\circ$ and $7^\circ$. 
This leads to a modest change of the efficiency $-0.2/+0.4$ and slight change of the magnetic field $\pm 0.2$~dex. 

\setlength{\tabcolsep}{10pt}
\renewcommand{\arraystretch}{1.2}
\begin{table}[!ht]
\caption{Log-likelihood values and test statistics for the full, FRONT and BACK-converted, and PSF0-3 data samples. For the conversion-type subsets, the full-data best-fitting $H_0$ and $H_1$ models were evaluated without refitting.}
\label{tab:front-back}
\centering
\begin{tabular}{lrrr}
\hline
Data sample & $\mathcal{L}_0$ & $\mathcal{L}_1$ & $TS$ \\
\hline
FRONT & $-680901.8$   & $-680893.4$   & $16.7$ \\
BACK  & $-598071.1$   & $-598059.8$   & $22.6$ \\
\hline
PSF0  & $-363683.5$   & $-363679.6$   & $7.9$  \\
PSF1  & $-391739.9$   & $-391739.5$   & $0.6$  \\
PSF2  & $-390269.9$   & $-390267.0$   & $5.9$  \\
PSF3  & $-389094.6$   & $-389089.2$   & $10.6$ \\
\hline
Full  & $-1027889.6$ & $-1027870.9$ & $37.4$ \\
\hline
\end{tabular}
\end{table}

\setlength{\tabcolsep}{13pt}
\renewcommand{\arraystretch}{1.3}
\begin{table}[!ht]
\caption{Systematic uncertainties associated with the nominal values for $\varepsilon_\mathrm{eff}$ and $B_\mathrm{rms}$ including instrumental effects (A$_{\mathrm{eff}}$ and PSF bracketing, for fixed $\ell_C=10~\mathrm{kpc}$) and theoretical uncertainties.}
\label{tab:systematics}
\centering
\begin{tabular}{lll}
\hline
      & $\varepsilon_\mathrm{eff}$ & $\log_{10}(B_\mathrm{rms} [\mathrm{G}])$\\ 
 Nominal values  & $0.97\pm 0.11$ & $-14.8^{+0.4}_{-0.9}$ \\
 \hline \hline
$A_{\mathrm{eff}}$ bracketing  & $\pm$ 0.02  & $<0.1$ \\ 
PSF($+5\%$/$-5~\%$)            & $-0.78/+4.54$    & $+0.8/-0.6$ \\ \hline
Instrumental                   & $-0.80/+4.56$ & $-0.6 / +0.8$\\
\hline
EBL model                      & $-0.02/+0.07$   & $<0.1$   \\ 
TeV injection                  &  $\pm 0.4$ &  $<0.1$   \\ 
$t_{dc}$ (low/high)            & $-0.2/+0.0$ &  $-1.4/+0.0$ \\ 
$\theta_\mathrm{jet}$ (large/small)         & $-0.2/+0.4$ & $\pm 0.2$ \\
\hline
Theoretical                    & $-0.82/+0.87$ & $-1.6/+0.2$\\
\end{tabular}
\end{table}

\setlength{\tabcolsep}{15pt}
\renewcommand{\arraystretch}{1.3}
\begin{table}[!ht]
\caption{Efficiencies $\varepsilon_{\mathrm{eff}}$ obtained for different EBL models included in ELMAG. The values were calculated for the best-fit $B_{\mathrm{rms}}$ and $\ell_{\mathrm{C}}$ parameters.}
\label{tab:EBL_eff}
\centering
\begin{tabular}{ll}
\hline
EBL model & $\varepsilon_{\mathrm{eff}}$ \\ \hline \hline
Finke \tablefootmark{ a}         &    1.04 $\pm$ 0.13  \\
Franceschini \tablefootmark{b} &    0.99 $\pm$ 0.12  \\
Dominguez \tablefootmark{c}    &    1.01 $\pm$ 0.12  \\
Gilmore \tablefootmark{d}      &    0.97 $\pm$ 0.11  \\
Kneiske$_{\mathrm{bf}}$ \tablefootmark{e} & 1.00 $\pm$ 0.12 \\
Kneiske$_{\mathrm{ll}}$ \tablefootmark{f} & 0.95 $\pm$ 0.11 \\
\end{tabular}
\tablefoot{
  \tablefoottext{a}{\citet{2010ApJ...712..238F}}
  \tablefoottext{b}{\citet{2008A&A...487..837F}}
  \tablefoottext{c}{\citet{2011MNRAS.410.2556D}}
  \tablefoottext{d}{\citet{2012MNRAS.422.3189G}}
  \tablefoottext{e}{\citet{Kneiske_2004}}
  \tablefoottext{f}{\citet{2010A&A...515A..19K}}
}
\end{table}

\section{Summary and Discussion}
\label{sec:conclusions}

The results of the analysis presented here can be summarized with a list of findings:
\begin{enumerate}
    \item Detection of extended emission at energies $>100$~MeV around Mkn~501 ($z=0.034$).
    \item The best-fitting model is the combination of a point-like source and a faint ( $\approx 6~\%$ of the total photon count of both sources) and extended (FWHM $\approx 1^\circ$) halo.  
    \item The halo emission is consistent with a pair-cascade origin for a magnetic field characterised by $B_\mathrm{rms}=1.5^{+2.0}_{-1.3}~\mathrm{fG}$  and a coherence length of $\ell_C=10 \mathbf{^{+490}_{-5}} ~\mathrm{kpc}$ (statistical uncertainties only).
    \item The statistical significance of the findings is estimated to be $5.0~\sigma$, including the look-elsewhere effect. 
    \item The best-fitting cascade flux is comparable  ($\varepsilon_\mathrm{eff}=0.97\pm0.11_\mathrm{stat}$, statistical uncertainties only) to that anticipated for the assumed source injection power ($\varepsilon_\mathrm{eff}=1$).
    \item The observed halo flux is consistent with the source being active for $t_{dc}\ge \num{45000}$~years (at 95~\% confidence level).
    \item Considering systematic uncertainties, the IGMF strength is estimated to be between 0.4~fG to 10 fG, for fixed $\ell_\mathrm{C} = 10~\mathrm{kpc}$.
\end{enumerate}

\subsection{Previous observations of Mkn~501}
Observations of  Mkn~501  have been previously considered  in the context of pair haloes: the non-detection of such a halo has been used to estimate a lower limit of
$B_\mathrm{rms} \ge 10^{-20}$ G for $\ell_C = 1$ kpc
\citet{2012ApJ...744L...7T}. The peculiar hard energy spectrum and delay of GeV emission during a flare of Mkn~501 in 2009
have been argued to be explainable by
cascade emission in a magnetic field of strength $10^{-17}~\mathrm{G}$ to
$10^{-16}~\mathrm{G}$ for a coherence length of $\ell_C \gg 1~\mathrm{Mpc}$ \citep{2012A&A...541A..31N}. Lastly, \citet{2018cosp...42E.606C} performed a likelihood ratio similar to the analysis carried out here of 12 stacked blazar sources, including Mkn~501, finding a $\approx 2~\sigma$ hint at $B_\mathrm{rms} \approx  10^{-15}$ G for $\ell_C=1~\mathrm{Mpc}$.

\subsection{Comparison with existing bounds}
In Fig.~\ref{fig:IGMF-par}, we present the accessible IGMF parameter space, including existing limits obtained with different methods.
This includes observation of Faraday rotation measure which is sensitive to magnetic fields irrespective of the epoch of its generation \citep{2016PhRvL.116s1302P,2022MNRAS.515..256P} and model-dependent limits on primordial magnetic fields obtained from CMB observations ranging from \SIrange{10}{1000}{\pico G}. \citep{2019PhRvL.123b1301J,2014PhRvD..89d3523T,2016A&A...594A..19P}.

 The first observational bound using \textit{Fermi}-LAT energy spectra of  Lac-type objects (1ES~0229+200 at $z=0.140$, 
 1ES~0347-121 at $z=0.188$, and 1ES~1101-232 at $z=0.186$)  set a lower-limit on the magnetic field  $\approx 10^{-16}~\mathrm{G}$ in the intervening medium \citet{2010Sci...328...73N}.  Similar lower limits have been obtained in the past with more refined methods 
\citep[see e.g.][]{2011ApJ...727L...4D,2023A&A...670A.145A, 2023ApJ...950L..16A,2024ApJ...963..135T,2025arXiv250622285B}. Some of these results
do not agree among each other; see e.g. \citet{2025arXiv250622285B} for a weaker bound in comparison to
\citet{2023ApJ...950L..16A} due to differences in the simulations and finally, \citet{2014ApJ...796...18A} demonstrates that a similar data set is consistent with a vanishing IGMF. In studies focussed on ground-based gamma-ray observations like VERITAS \citep{2017ApJ...835..288A} and H.E.S.S. \citep{2014A&A...562A.145H}, 
optimistic assumptions not supported by observations on the injection spectrum are used to obtain limits on the IGMF which subsequently are considerably larger ($\ge 10^{-14}$~G) than the ones obtained from the abovementioned studies.

A more generic and model-independent search for a halo-like emission in high Galactic latitude sources has been carried out by the Fermi collaboration \citep{2018ApJS..237...32A}, where 7.5 years of data above 1 GeV were analysed and no evidence for extension was found.  The apparent discrepancy is a result of the assumption that the extension is energy-independent while the cascade emission is mainly observed at several GeV. We have carried out an analysis in this data set using fixed-extension templates and found only a marginal value for the test statistic ($TS_\mathrm{ext}=2.4$), consistent with the result presented in \citet{2018ApJS..237...32A}.
 
 For a  more meaningful comparison with the results presented in this paper, we consider recent bounds by \citet{2023ApJ...950L..16A} that have been established using methods that are very similar to the ones applied here\footnote{Note, their simulation of the cascades is done using CRPROPA and implements a cell-like magnetic field structure}.  
 Among the five sources used in that analysis, we consider the two sources with the highest (1ES~1101-232 at $z=0.186$) and lowest (PKS~0548-322 at $z=0.069$) isotropic electron-injection luminosity which leads to the highest and respectively lowest limits on the B-field (see Fig.~\ref{fig:IGMF-par}; the values shown have been taken from Fig.~7 of their paper for $t_{dc}=\SI{e7}{\year}$)\footnote{Note, the confidence level of the limits is actually 90~\% and not as claimed by \citet{2023ApJ...950L..16A} at the 95~\% level}. 
 
 The electron-injection luminosity $L_{iso,e}$ relates the power of the source to the power that is injected via pair-production processes and is given by
 
 \begin{ceqn}
 \begin{equation}
     L_{iso,e} = 4\pi D_L^2 \int\limits_{E_l}^{E_h} \mathrm{d}E~ 
     E \frac{\mathrm{d}N}{\mathrm{d}E} \left( 1-e^{-\tau_{\gamma\gamma}(E,z)}\right),
     \label{eqn:liso}
\end{equation}
\end{ceqn}

with the differential photon flux $\mathrm{d}N/\mathrm{d}E$ given by Eqn.~\ref{eq:inj_power}, the luminosity distance $D_L(z)$, calculated using Planck18 cosmology \citep{2020A&A...641A...6P}, and $\tau_{\gamma\gamma}(E,z)$ the optical depth at observed photon energy $E$ and red-shift $z$ from \citet{2010ApJ...712..238F}. The integration range is chosen to be $E_l=\SI{0.1}{\tera\eV}$ and $E_h=\SI{31.6}{\tera\eV}$ as introduced in \citet{2023ApJ...950L..16A}.
For Mkn~501, we find $L_{iso,e}(\mathrm{Mkn~501})=\SI{4.4e43}{\erg\per\second}$. 
For the nearby source PKS~0548-322, the best-fitting injection spectrum listed in Table~4 of \citet{2023ApJ...950L..16A} for $t_{dc}=\SI{e7}{\year}$ and $B_{rms}=\SI{3.16e-16}{G}$ the resulting 
$L_{iso,e}(\mathrm{PKS~0548-322})=\SI{2.6e44}{\erg\per\second}$ is about a factor of five larger than $L_{iso,e}(\mathrm{Mkn~501})$. Finally, for 1ES~1101-232, the best-fitting spectrum at $B_{rms}=\SI{3.16e-14}{G}$ results in an estimate of $L_{iso,e}(\mathrm{1ES~1101-232})=\SI{1.7e45}{\erg\per\second}$ that is about six times more than PKS~0548-322.

With increasing energy density of pairs in the intervening medium, the energy-loss rate of electrons and positrons related to plasma-heating processes 
is expected to increase as suggested by \citet{2012ApJ...752...22B}. In the case of Mkn~501, we find an efficiency of $\varepsilon_\mathrm{eff}\approx 1$ while for more luminous sources, this efficiency could be $\varepsilon_\mathrm{eff}\ll 1$ because of non-radiative energy-loss mechanisms dominating over inverse-Compton losses. 
With sub-dominant radiative processes, the cascade emission is quenched and the lower bounds derived on the magnetic field would have to be corrected for plasma heating processes. In a more recent theoretical study \citep{2022ApJ...929...67A}, the authors found that the presence of sizeable IGMF suppresses non-radiative energy losses.  
In this scenario, the absence of cascade emission implies either a lower bound (well above the favoured IGMF found here) or an upper bound on the IGMF (well below the favoured IGMF).

Besides the difference in the integrated luminosity given by Eqn.~\ref{eqn:liso}, the energy at which the integrand of that function peaks is close to 10~TeV for 1ES1101-232 while for Mkn~501, it is at $\approx 1.5~\mathrm{TeV}$. As a consequence of this, the cascade for 1ES~1101-232 develops within $\approx 10$~Mpc while for Mkn~501, the injected pairs sample more evenly the IGMF along the line of sight. This fact may offer an explanation for the apparent inconsistency: While the pairs injected from Mkn~501 sample the IGMF in the void volume, most
of the pairs injected by 1ES1101-232 will be affected by the filamentary magnetic fields ($\mathcal{O}(10~\mathrm{nG})$ in the environment of 1ES1101-232 \citep{2025A&A...693A.208C}.

Additionally, we include the upper limits obtained by \citet{2018ApJ...868...87B} resulting from the non-observation of gamma-ray bow-ties from misaligned AGN, which set an upper bound of $10^{-15}$G for Mpc-scale correlation lengths, well below the lower-bound reported for 1ES1101-232. This limit is shown as a solid black line in Fig. \ref{fig:IGMF-par}. 

\subsection{Comparison with model predictions}
\label{subsec:comparison}
\begin{figure}
   \centering
   \includegraphics[trim=0 0 0 0,clip,scale=0.46]{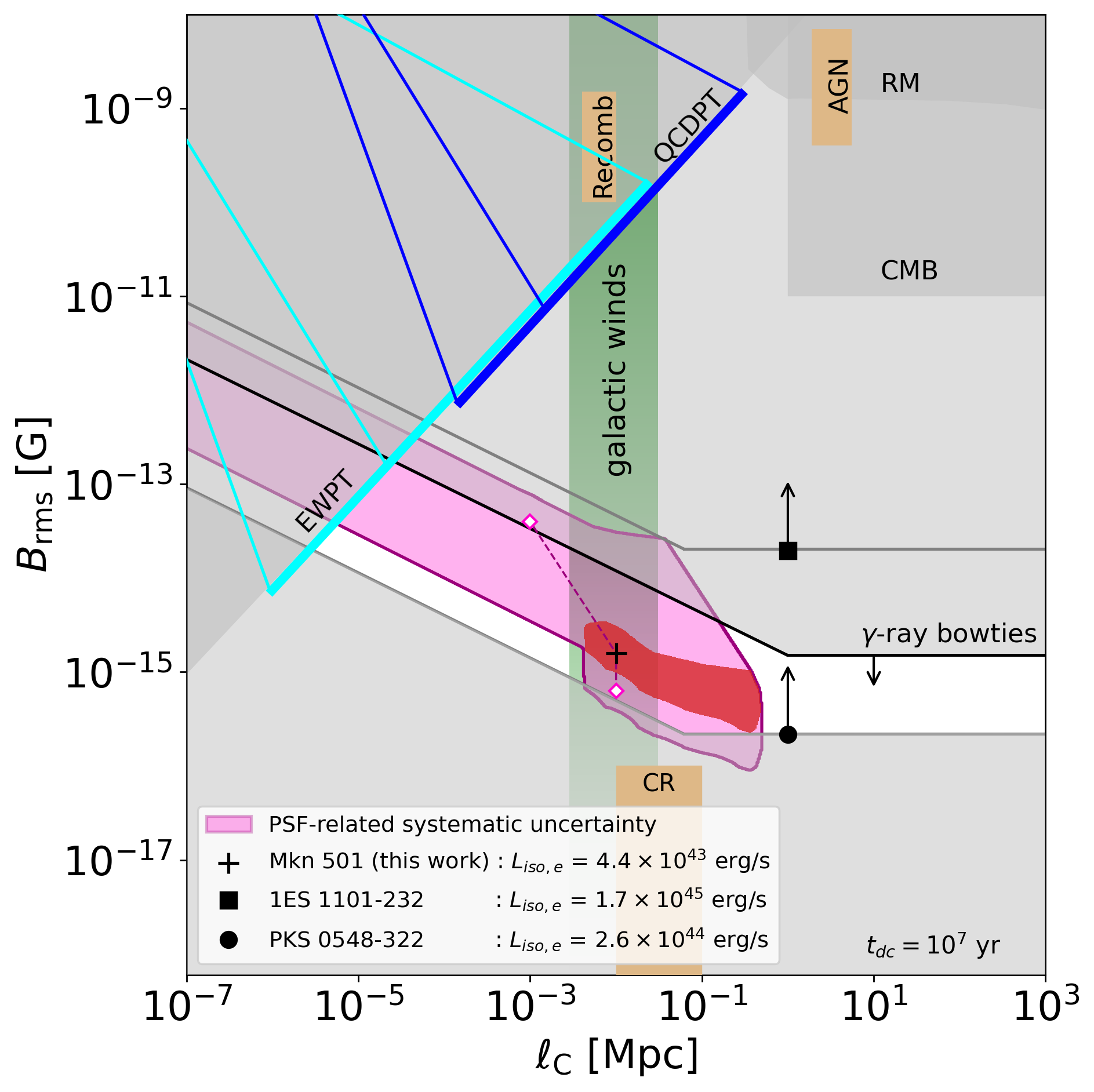}
   \caption{Parameter space  of the  magnetic field $B_\mathrm{rms}$ and coherence length $\ell_C$ in the intergalactic medium: Grey shaded regions are excluded, while the orange and green regions are representative predictions from astrophysical processes. The evolved fields of primordial origin are expected to exist within the white region with representative tracks for magnetogenesis during the QCD and electro-weak phase transitions indicated in blue and cyan.  The plus symbol represents the best-fitting model for Mkn 501 from this work, with the 68 percent contour shown in red (statistical uncertainty only), and the systematic shift presented in pink. See the text in Section~\ref{subsec:comparison} for further details. 
   }
   \label{fig:IGMF-par}
\end{figure}

The best-fitting parameters for $B_\mathrm{rms}$ and
$\ell_C$ are specific to the line of sight towards
Mkn~501. The injection of electrons occurs along the mean-free path of the photons which ranges from $\approx 1~\mathrm{Mpc}$ at the highest photon energies to several hundred Mpc at TeV energies. The magnetic field affecting the electron/positrons along the line of sight most likely changes from $\mu$G fields close to the host galaxy to the weaker field within the low-density region of voids. The magnetic field present in the void could be 
predominantly of primordial origin related to processes taking place in the early universe including electroweak and QCD symmetry-breaking \citep[for reviews, see e.g.][]{KANDUS20111,2013A&ARv..21...62D}.  The generated magnetic field evolves dynamically until recombination \citep{2004PhRvD..70l3003B} (see tracks in Fig.~\ref{fig:IGMF-par})  and is subject
to amplification via gravitational instabilities and modifications by astrophysical processes. A robust prediction of  primordial magnetogenesis
is a bound related to the largest eddy size with a turn-over time $\ell_C/v_A\le t_0$ at the time of recombination $t_0$,
 with $v_A$ the Alfvén velocity such that 
$B  \le 10^{-8}~\mathrm{G} (\ell_C/\mathrm{Mpc})$. This bound is indicated by the diagonal cyan and blue-coloured band in Fig.~\ref{fig:IGMF-par}.

Besides the primordial magnetogenesis, there are several astrophysical processes that potentially contribute to the magnetic field present in the
voids (orange-coloured regions in Fig.~\ref{fig:IGMF-par}):
\begin{description}
\item{\textbf{Recombination:}} The relaxation processes during the epoch of recombination will drive turbulences which subsequently can lead to magnetic field
generation. A recent radiative hydrodynamic  simulation resolves a large dynamic range of turbulences. In case of saturation of the
magnetic fields, large magnetic field ($\approx$~nG) generation can be expected \citep{2025arXiv250421082C}. 
\item[\textbf{Cosmic rays}] accelerated by supernova explosions of the first stars will generate magnetic fields $\approx 10^{-17\ldots -16}$~G at $\ell_{C}\approx 100~\mathrm{kpc}$ during the phase of re-ionization \citep{2011ApJ...729...73M}. This prediction is particularly robust under changes of the parameters of the model. 
\item[\textbf{AGN feedback}:] The relativistic outflows of active galactic nuclei will fill bubbles with a comoving radius  of several Mpc which are filled with 
a magnetic field up to nG strength \citep{2001ApJ...556..619F, 2022A&A...660A..80B}. 
\item[\textbf{Galactic winds}]  release a magnetized plasma into the IGMF \citep[see e.g.][]{2006MNRAS.370..319B}, such that the resulting magnetic fields 
reach values as high as $10^{-12}$~G with large filling-factors in the environment of galaxies. See however, the discussion 
in \citet{2012ApJ...752...22B} related to the velocity of the escaping winds which would limit the filling factor of this contribution.  

\end{description}

The pink shaded region in Fig.~\ref{fig:IGMF-par} tracks the nominal uncertainty contour towards the best-fit PSF bracketing points presented in Appendix~\ref{app:PSF-bracketing}, since the 68 percent contour of the "Wide PSF" does not close for small coherence lengths, the uncertainty region is extended accordingly.

The results presented here are consistent with both a primordial and an astrophysical origin. Even if primordial magnetic fields remain unperturbed in voids, a contribution from astrophysical processes appears unavoidable. Most of the aforementioned astrophysical processes tend to generate fields at the nG-strength which are not excluded by the observations presented here. Even though the observations indicate fG fields, the measurement would not be sensitive to nG fields with a filling factor well below unity \citep{2011ApJ...727L...4D}.

\begin{acknowledgements}
MS acknowledges funding by the Deutsche Forschungsgemeinschaft (DFG, German Research Foundation) under Germany’s Excellence Strategy–EXC 2121 “Quantum Universe” – 390833306. The authors thank Marcus Brüggen, Manuel Meyer and Günter Sigl for helpful comments. 
\end{acknowledgements}

%
\bibliographystyle{bibtex/aa} 
\bibliography{bibtex/ref} 
%

\begin{appendix}
\section{Simulating a source with finite duty cycle}
\label{app:duty-cycle}

The source at a distance \(R\) injects a differential rate of photons,
such that the energy and time-dependent part can be factorized:
\begin{ceqn}
\begin{equation}
\label{eq:ndot}
 \dot n_\mathrm{inj}(t,E) = N_0~g(t)~f(E),
\end{equation}
\end{ceqn} 
where \(N_0\) is the total number of photons, \(g(t)\) is the probability for a photon to pass through a sphere of radius \(R\) centred on the source in the interval \(t\) and \(t+\mathrm{d}t\), and \(f(E)\) is the probability to inject a photon of energy in the interval \(E\) and \(E+\mathrm{d}E\).

The simulation of the cascade process is used to estimate the resulting rate of photons of energy \(E_\gamma\) that pass a sphere of radius \(R\)
at time $t_\gamma$ with a time delay \(t'=t_\gamma-R/c\).

This way, we obtain the function \(G(t'>0,E_\gamma<E_0;E_0)\) by injecting instantaneously mono-energetic photons of energy \(E_0\) and registering the secondary photons in the interval 
$E_\gamma$ to $E_\gamma + \mathrm{d} E_\gamma$ with their individual time-delay between \(t'\) and \(t'+\mathrm{d}t'\).

The resulting photon rate at time \(t'\) follows from the convolution:
\begin{ceqn}
\begin{equation}
  \dot n(t', E_\gamma) = N_0\int\limits_{E_\gamma}^{\infty} \mathrm{d}E_0~f(E_0) 
  \int\limits_{-\infty}^{+\infty}\mathrm{d}t~ g(t)G(t'-t,E_\gamma;E_0),
\end{equation}
\end{ceqn} 
or in a more compact notation: 
\begin{ceqn}
\begin{equation}
  \dot n(t', E_\gamma) = N_0\int\limits_{E_\gamma}^{\infty} \mathrm{d}E_0~f(E_0)
  (g\ast G)(t').
\end{equation}
\end{ceqn} 

In the assumed scenario of a constant injection rate over the duty cycle time \(t_{dc}\) and starting at a time \(t=0\), the time-dependent part is simply given by 
\begin{ceqn}
\begin{equation}
 g(t) = \frac{1}{t_{dc}} H(t)H(t_{dc}-t).
\end{equation}
\end{ceqn}
In this case and assuming \(t'= t_{dc}\), we can simplify: 
\begin{ceqn}
\begin{equation}
\label{eq:a5}
 \dot n(t_{dc}, E_\gamma) = \frac{N_0}{t_{dc}} \int\limits_{E_\gamma}^{E_\mathrm{max}}\mathrm{d}E_0~f(E_0) 
 \int\limits_{0}^{t_{dc}}\mathrm{d}t~G(t_{dc}-t,E_\gamma;E_0).
\end{equation}
\end{ceqn} 

This is the approach chosen here to consider the effect
of the duty cycle \(t_{dc}\) on the observed cascade photon rate.

\begin{figure}
   \centering
   \includegraphics[trim=0 0 0 0,clip,scale=0.45]{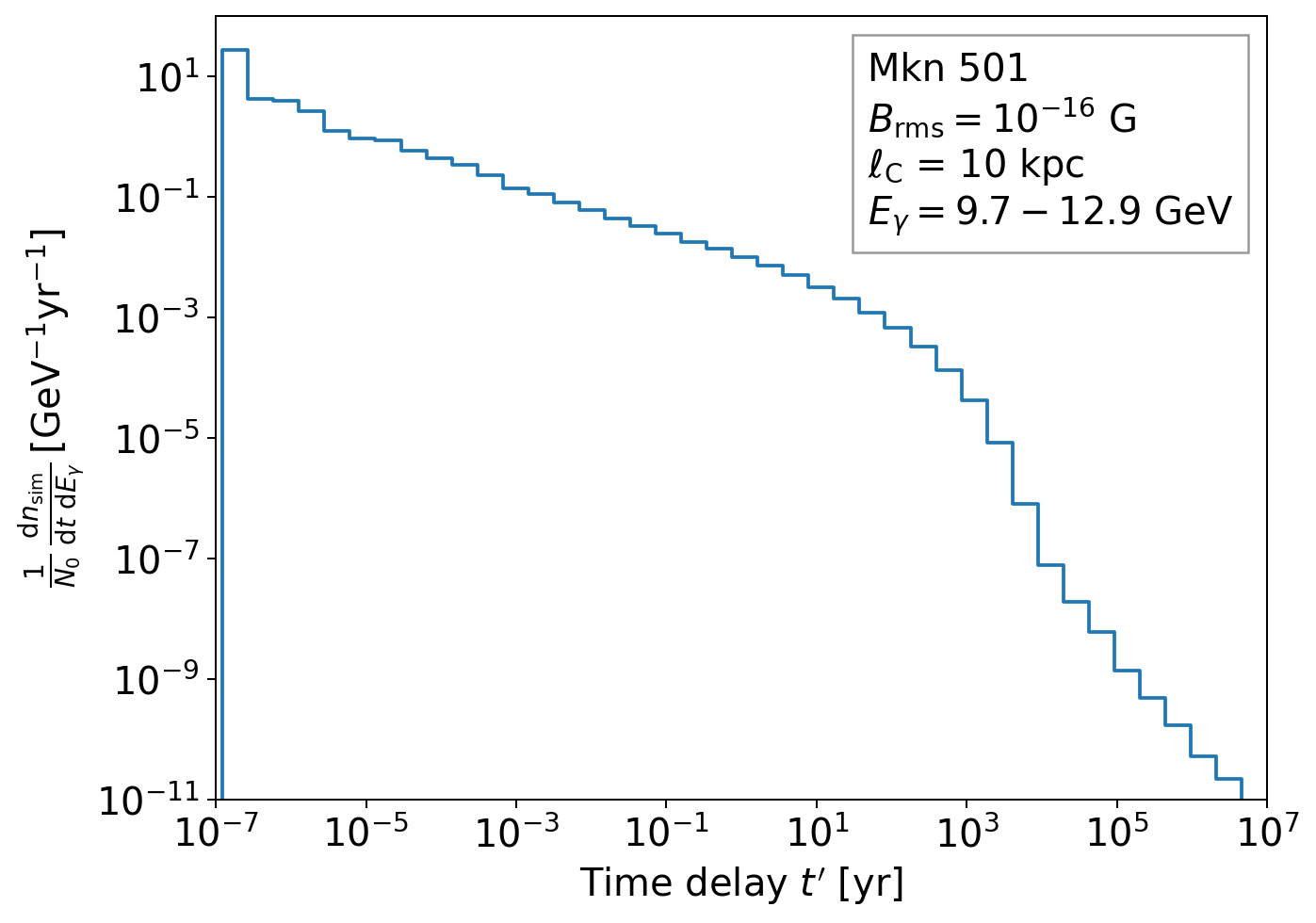}
   \caption{Time delay distribution of secondary photons resulting from the simulations assuming an instantaneous injection, for the parameters given in the legend. This distribution is represented by Eqn. \ref{eq:a6}.}
   \label{fig:simulation_appendix}
\end{figure}

As a demonstration of the simulation result, we show in Fig.~\ref{fig:simulation_appendix} the result of an instantaneous injection $g(t)=\delta(t)$, such that 
\begin{ceqn}
\begin{equation}
\label{eq:a6}
\frac{\mathrm{d}n_\mathrm{sim}(t',E_\gamma)}{\mathrm{d}t~\mathrm{d}E_\gamma} = 
N_0 \int\limits_{E_\gamma}^{E_\mathrm{max}} \mathrm{d}E_0~f(E_0)G(t',E_\gamma;E_0).
\end{equation}
\end{ceqn}

We evaluate
\begin{ceqn}
\begin{equation}
\dot n_\mathrm{sim}(t_{dc},E_\gamma)= \frac{1}{t_{dc}} \int\limits_{0}^{t_{dc}}\mathrm{d}t'~\frac{\mathrm{d}n_\mathrm{sim}}{\mathrm{d}t \mathrm{d}E_\gamma},
\end{equation}
\end{ceqn}
which equals Eqn.~\ref{eq:a5} (constant injection rate over time-scale $t_{dc})$  to compute the rate
for a constant injection 
over $t_{dc}$.  

\section{Time-delay simulations with ELMAG}
\label{app:time-delay}
The results presented in the main text assumed a maximum time-delay of $10^7$ years. 
Additionally, we consider template cubes that were accumulated for delay times from $10^3$ to $10^7$ years, in order to investigate the effect of different duty cycles of the source. 

\begin{figure}
   \centering
   \includegraphics[trim=0 0 0 0,clip,scale=0.42]{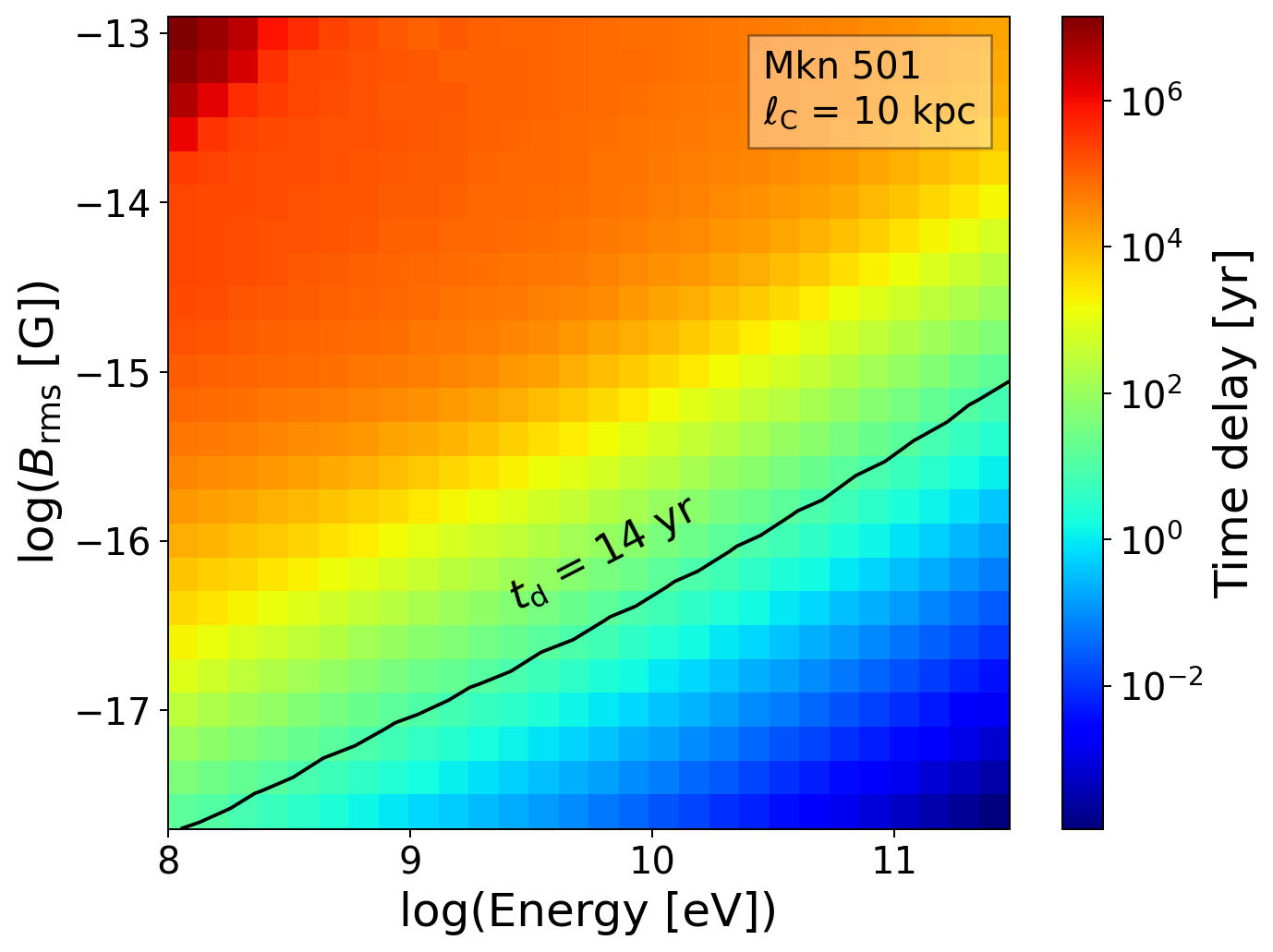}
   \caption{Median time delay of the secondary photons registered in ELMAG for different $B_\mathrm{rms}$ strengths and energy bins. 
   The solid line follows the iso-contour related to the time covered with $\it{Fermi}$-LAT observations.}
   \label{fig:mkn501-td-grid}
\end{figure}

The time-delay increases with decreasing energy and increasing magnetic field as expected (see e.g., Eqn.19 of ~\citet{2021Univ....7..223A}). This can be seen in Fig.~\ref{fig:mkn501-td-grid}, where the median time-delay of photons is shown with a colour scale in the plane of detected photon energy and $B_\mathrm{rms}$, for a fixed value of the coherence length $\ell_C=\SI{10}{\kilo\parsec}$.   

\subsection{Duty cycle of Mkn~501}
\label{sec:duty-cycle}
\begin{figure}
   \centering
   \includegraphics[trim=0 0 0 0,clip,scale=0.5]{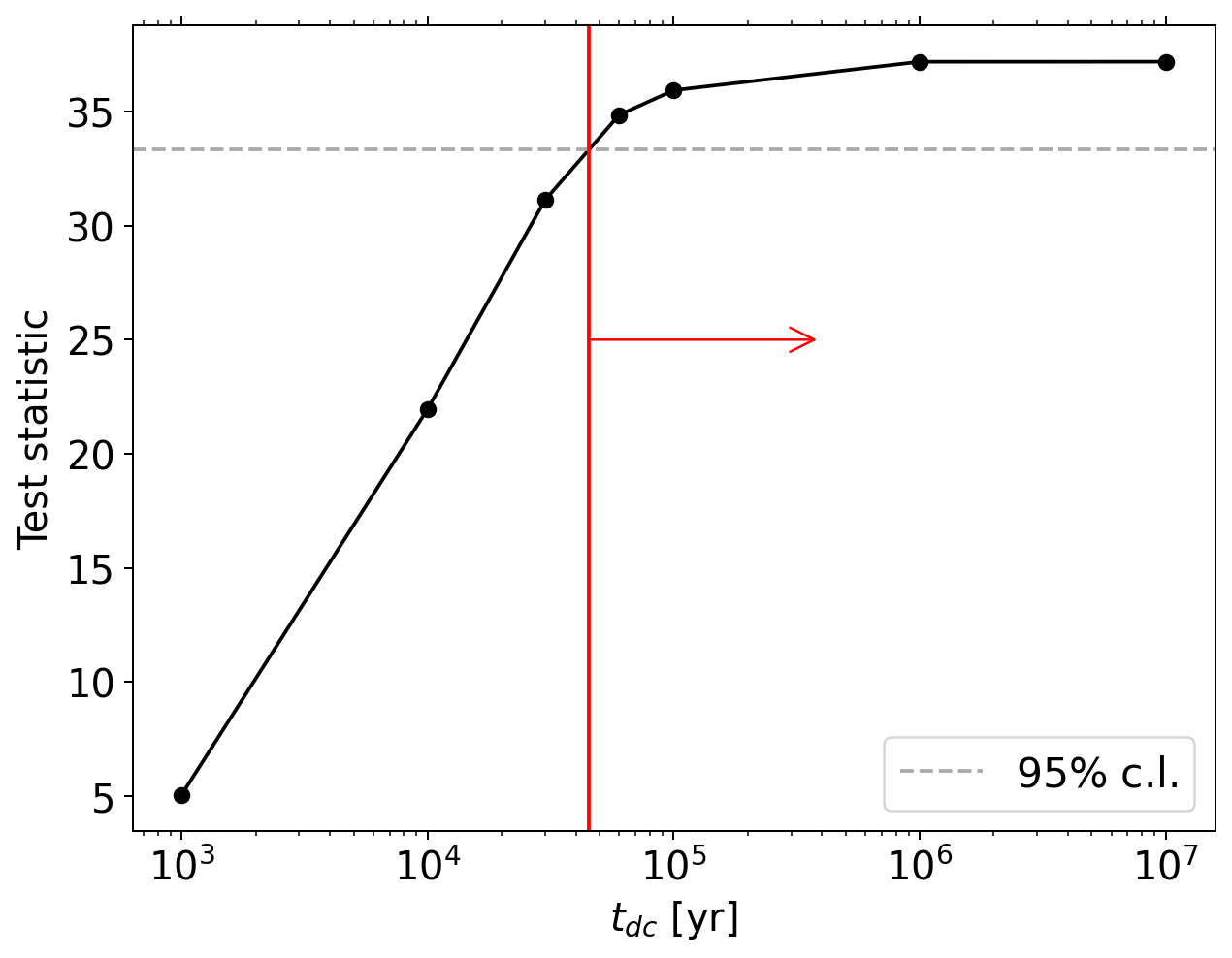}
   \caption{Test statistic profile obtained from our fit to each time delay bin. Dashed line shows the 95$\%$ confidence interval.}
   \label{fig:mkn501-td}
\end{figure}

 So far, the time during which the source is actively injecting TeV photons has been considered to be larger than $10^{7}$ years (in the following this time-scale of activity is called \textit{duty cycle} $t_{dc}$). The first detection of TeV $\gamma$-rays in 1995 \citep{1996ApJ...456L..83Q} sets a lower-limit on the duty cycle of $\approx 30$ years. To extend beyond the steady-state approach followed so far, the simulations have been extended to cover duty cycles of $\SIrange{e3}{e7}{years}$. For each duty cycle, the analysis is repeated for a fixed $\ell_C=\SI{10}{\kilo\parsec}$ and varying $B_\mathrm{rms}$ and efficiency $\varepsilon_\mathrm{eff}$.
 The resulting $TS$  for each value of $t_{dc}$ is shown in Fig.~\ref{fig:mkn501-td}. 
 At $t_{dc}\gtrsim 4.5\times 10^4$~years, the test statistics obtained before is recovered.  
 This result can be used to estimate a lower bound to the duty cycle of this source to be $t_{dc}>\num{45000}~\mathrm{yrs}$ with a confidence interval of $\approx \num{95}~\%$, interpolating the $TS$ values between the different simulations.

\section{PSF bracketing}
\label{app:PSF-bracketing}

The search for a faint cascade around Mkn~501 is particularly sensitive to the point-spread function (PSF). The point source and cascade contribute photons in overlapping regions of energy and angular space. A small PSF mismatch can therefore change both the detection significance and the inferred IGMF parameters. 
We investigate this by completing the full $24\times9$ IGMF scan reported in the main text, implementing the PSF bracketing introduced in \ref{sec:systematics}. The narrow and wide PSF realisations were generated by scaling the \texttt{SCORE} and \texttt{STAIL} parameters by 0.95 and 1.05, respectively, providing controlled $-5\% / +5\%$ angular re-scaling of the nominal double-King profile.

\begin{figure}
   \centering
   \includegraphics[trim=0 0 0 0,clip,scale=0.5]{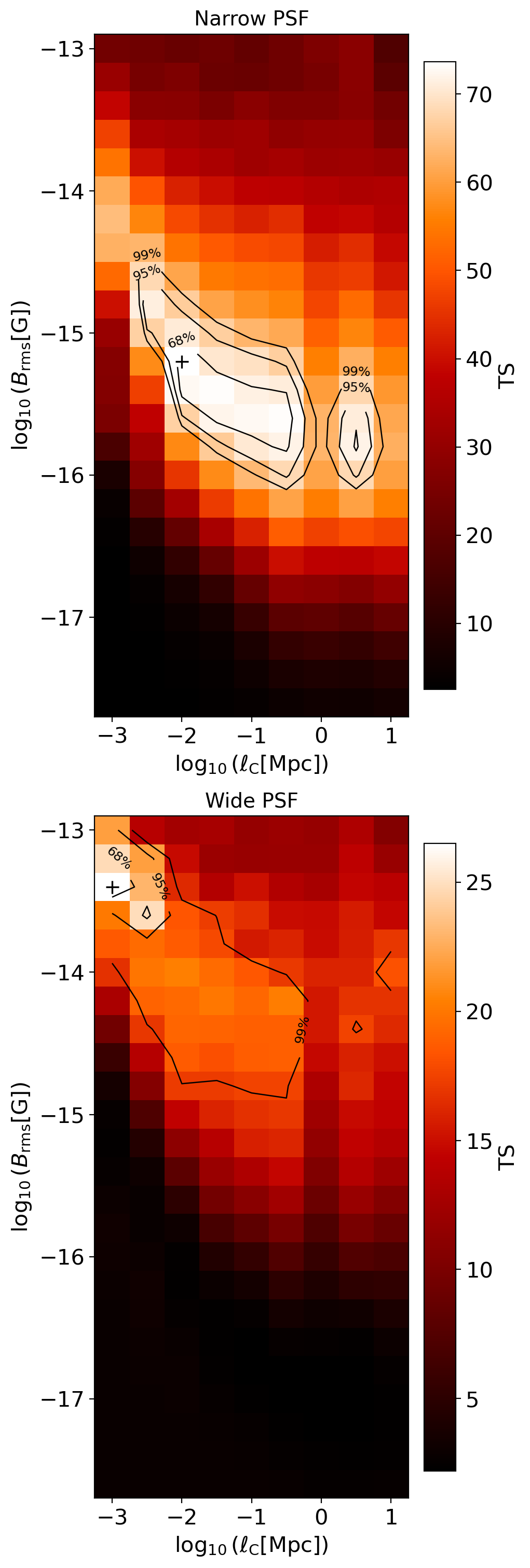}
   \caption{Test statistic grids obtained from the analysis under the narrow and wide PSF assumptions.}
   \label{fig:mkn501-PSF-bracket}
\end{figure}

Figure~\ref{fig:mkn501-PSF-bracket} shows the Test Statistic grid obtained from the bracketing analysis. The best-fit parameters, maximum TS and null likelihood $\mathcal{L}_0$ are summarized in Table~\ref{tab:psf-bracket}. The narrow response yields a maximum of TS$_\mathrm{narrow}=73.7$ for IGMF parameters $B_\mathrm{rms} = 0.63$ fG and $\ell_\mathrm{C}=10$ kpc, strongly in agreement with the nominal result. 
Whereas the wide realization results in a lower maximum TS$_\mathrm{wide}=26.8$, for a stronger IGMF of $B_\mathrm{rms} = 39.8$ fG and shorter $\ell_\mathrm{C}=1$ kpc. This behaviour is expected from the competing point-source tails and extended components. The preferred cascade templates are correspondingly shifted towards morphologies produced by larger magnetic deflections.

\setlength{\tabcolsep}{7pt}
\renewcommand{\arraystretch}{1.3}
\begin{table}[t]
  \centering
  \caption{IGMF parameters, maximum TS values and corresponding null hypothesis likelihoods obtained for the nominal as well as the two bracketing PSF models.}
  \label{tab:psf-bracket}
  \begin{tabular}{lcccc}
    \hline
    Response & $B_{\rm rms}$ (fG) & $\ell_C$ (kpc)
             & TS$_\mathrm{max}$ & $\mathcal{L}_0$\\
    \hline
    Nominal & $1.50$  & $10$ & $37.4$ & $\num{-1027889.69}$\\
    Narrow  & $0.63$ & $10$ & $73.7$ & $\num{-1027948.10}$\\
    Wide    & $39.8$  & $1$ & $26.8$ & $\num{-1027893.32}$\\
    \hline
  \end{tabular}
\end{table}

These results showcase the sensitivity of our analysis to the specific instrument response function assumed, however within this controlled evaluation, the cascade signal remains significant.

\subsection{Pulsar-based PSF model}
In recent work by \cite{2026arXiv260214764B}, an alternative point spread function model is derived using phase-subtracted pulsar data. The authors claim that the cascade detection presented here is not observed when their instrument response model is used in the likelihood analysis, however they test a single configuration of the $B_{\rm rms}$ and $\ell_C$ parameters.

\begin{figure}
   \centering
   \includegraphics[trim=0 0 0 0,clip,scale=0.63]{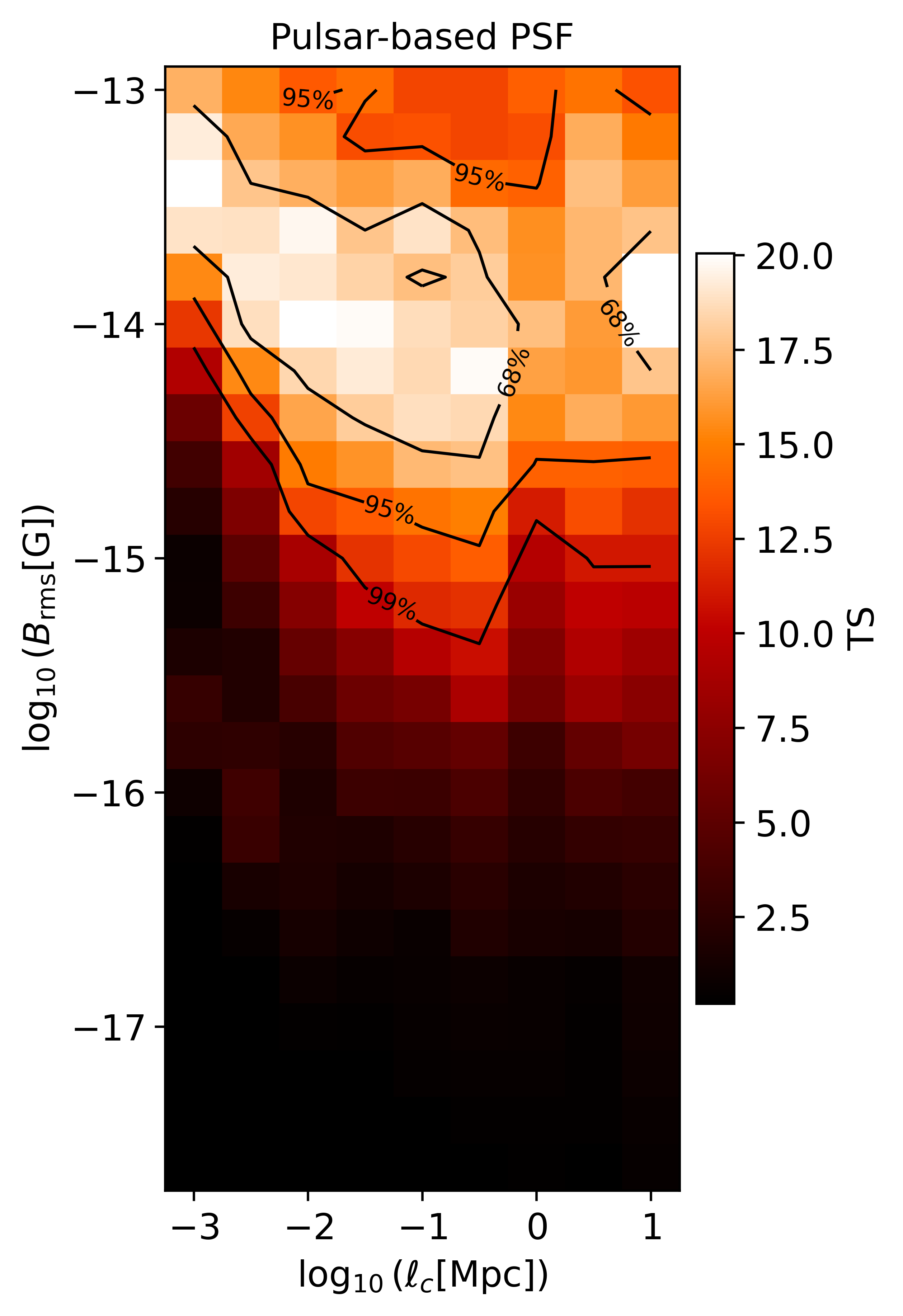}
   \caption{Test statistic grid from the analysis assuming the pulsar-based PSF model.}
   \label{fig:mkn501-PSF-pulsar}
\end{figure}

To properly corroborate this claim, we performed the analysis over the full $B_{\rm rms}$, $\ell_C$ grid using their model, shown in Fig~\ref{fig:mkn501-PSF-pulsar}. 
We find that the cascade signal does not vanish, but rather the preferred region shifts towards stronger magnetic fields, as would be expected when using a broader PSF. This effect is similar to that observed from our PSF bracketing analysis.

To further evaluate their model, we performed a likelihood comparison between the nominal and the pulsar-based models, analysing the ROIs of high-galactic-latitude sources to determine whether the proposed model represent an improvement on the fit, the results are summarized in Table~\ref{tab:psf_loglike_comparison}. We found that for all of the 10 tested sources, the nominal PSF model is largely preferred in terms of likelihood.

\begin{table}[!h]
\centering
\caption{Energy-dependent log-likelihood differences,
$\Delta\log\mathcal{L} =
\log\mathcal{L}_{\mathrm{Pulsar}} -
\log\mathcal{L}_{\mathrm{Nominal}}$.
Negative values favour the nominal PSF model.}
\label{tab:psf_loglike_comparison}

\footnotesize
\setlength{\tabcolsep}{2pt}
\renewcommand{\arraystretch}{1.15}

\resizebox{\columnwidth}{!}{%
\begin{tabular}{@{} l
    S[table-format=-3.2]
    S[table-format=-3.2]
    S[table-format=-2.2]
    S[table-format=-2.2]
    S[table-format=-1.2]
    | S[table-format=-3.2] @{}}
\toprule
& \multicolumn{5}{c}{Energy range (GeV)} & \\
\midrule
Source
& {\shortstack{0.10 --\\0.31}}
& {\shortstack{0.31 --\\0.98}}
& {\shortstack{0.98 --\\3.09}}
& {\shortstack{3.09 --\\9.70}}
& {\shortstack{9.70 --\\300}}
& {\shortstack{Full\\range}} \\
\midrule
Mkn 421
& -21.85 & -7.99 & -11.91 & -11.19 & -0.46 & -53.40 \\
3C 279
& -61.02 & -19.34 & -4.74 & -0.62 & 2.85 & -82.88 \\
PKS 2155-304
& -7.64 & -4.28 & -0.29 & -1.99 & 1.83 & -12.36 \\
PKS 1510-089
& -64.36 & -20.65 & -0.91 & -10.06 & 0.40 & -95.57 \\
Mkn 501
& -28.67 & -7.56 & -8.75 & -2.73 & 0.14 & -47.57 \\
CTA 102
& -71.04 & -44.32 & -13.35 & -12.38 & -1.64 & -142.73 \\
3C 454.3
& -181.76 & -62.50 & -27.76 & -15.88 & 0.50 & -287.40 \\
4C +55.17
& -11.26 & -4.36 & -0.33 & -4.66 & -0.70 & -21.31 \\
1H 1013+498
& -3.02 & -1.14 & 1.53 & -2.94 & 0.53 & -5.05 \\
PKS 1424+240
& -3.44 & -3.32 & 8.45 & -4.59 & -0.90 & -3.79 \\
\midrule
\bfseries Combined
& -454.05 & -175.47 & -58.06 & -67.03 & 2.55 & -752.07 \\
\bottomrule
\end{tabular}%
}
\end{table}

This leads us to conclude that the model proposed in \cite{2026arXiv260214764B} does not represent an improvement over the nominal instrument response, and furthermore, we show that the cascade signal remains significant even if this alternative model is assumed.

\end{appendix}

\label{LastPage}
\end{document}